\documentclass[aps,superscriptaddress,nofootinbib,floatfix,epfs,preprintnumbers]{revtex4}
\usepackage{amsfonts,amscd,amsmath,amssymb,graphicx,color,float,relsize}
\usepackage[section]{placeins}
\def\ep{\text{e}}
\def\g{\mathsf{g}}
\def\oh{\frac{1}{2}}
\def\s{\mathsf{s}}
\def\m{\mathsf{m}}
\def\k{\mathsf{k}}
\def\n{\mathsf{n}}

\def\km{-\frac{1}{4}\ep^{\frac{1}{4}}}

\def\rq{r_q}
\def\rv{r_v}

\def\Qqb{\text{\tiny Q}\bar{\text{\tiny q}}}

\def\QQb{\text{\tiny Q}\bar{\text{\tiny Q}}}
\def\Qqq{\text{\tiny Qqq}}
\def\QQq{\text{\tiny QQq}}
\def\QQ{\text{\tiny QQ}}
\def\Se{\text{\tiny (S)}}
\def\Me{\text{\tiny (M)}}
\def\Le{\text{\tiny (L)}}
\def\3Q{3\text{\tiny Q}}
\def\qQb{\text{\tiny q}\bar{\text{\tiny Q}}}
\def\3Q{3\text{\tiny Q}}
\def\qqq{3\text{\tiny q}}

\def\l31{\ell_{\3Q}^{(1)}}
\def\vp{v_{\shortparallel}}
\def\vet{v_{\mathsmaller{\vartriangle}}}
\def\vit{v_{\mathsmaller{\blacktriangle}}}
\def\vm{v_\text{\tiny M}}
\def\vl{v_\text{\tiny L}}
\newcommand{\xdownarrow}[1]{%
  {\left\downarrow\vbox to #1{}\right.\kern-\nulldelimiterspace}}

\begin{document}
\preprint{LMU-ASC 47/20}
\title{Remarks on Static Three-Quark Potentials, String Breaking and Gauge/String Duality}
\author{Oleg Andreev}
 \affiliation{L.D. Landau Institute for Theoretical Physics, Kosygina 2, 119334 Moscow, Russia}
\affiliation{V.A. Steklov Mathematical Institute, Gubkina 8, 119991, Moscow, Russia}
\affiliation{Arnold Sommerfeld Center for Theoretical Physics, LMU-M\"unchen, Theresienstrasse 37, 80333 M\"unchen, Germany}
\begin{abstract} 
Making use of the gauge/string duality, it is possible to study some aspects of the string breaking phenomenon in the three quark system. Our results point out that the string breaking distance is not universal and depends on quark geometry. The estimates of the ratio of the string breaking distance in the three quark system to that in the quark-antiquark system would range approximately from $\frac{2}{3}$ to $1$. In addition, it is shown that there are special geometries which allow more than one breaking distance. 
 
\end{abstract}
\maketitle
\section{Introduction}
\renewcommand{\theequation}{1.\arabic{equation}}
\setcounter{equation}{0}

Since the $J/\Psi$ discovery in 1974, there remains a serious puzzle concerning triply heavy baryons \cite{bj}. On the theoretical side, the challenge is to explain the structure and properties of such baryons, and thus reach the level of knowledge similar to that of charmonium and bottomonium. It might be expected that, based on the success with the quarkonium spectroscopy, the potential (quark) models would be useful in doing so \cite{richard}. At that stage, one would expect to gain important insights into understanding how baryons are formed from quarks and gluons.

\begin{figure}[htbp]
\centering
\includegraphics[width=3.5cm]{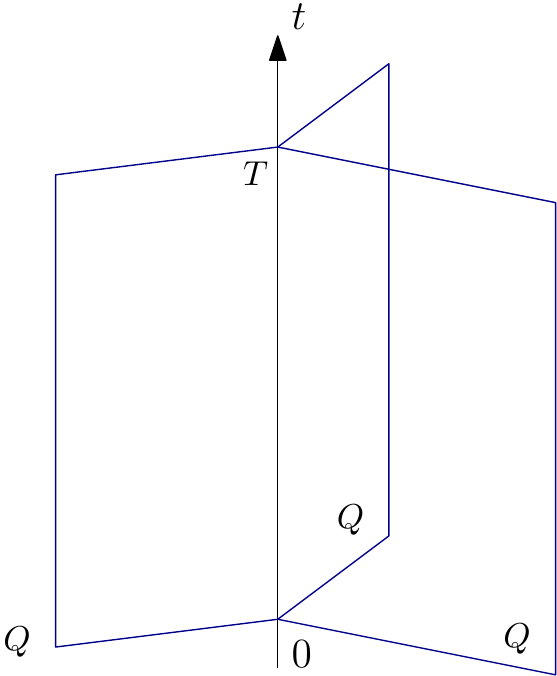}
\caption{{\small A baryonic Wilson loop in $SU(3)$ gauge theory. A three-quark state is generated at $t=0$ and then annihilated at $t=T$.}}
\label{Wl}
\end{figure}

A static three-quark potential is one of the most important inputs to the potential models and also a key to understanding the phenomenon of quark confinement in baryons. The potential as a non-perturbative object has been studied in the context of lattice gauge theory \cite{bali}.\footnote{For more recent developments, see \cite{3Q-lattice}.} It is determined from the expectation value of a baryonic Wilson loop. Such a loop is defined in a gauge-invariant manner as $W_{\3Q} =\frac{1}{3!}\epsilon_{abc}\epsilon_{a'b'c'} U_1^{aa'} U_2^{bb'} U_3^{cc'}$, with $U_i$ the path-ordered exponents along the paths shown in Figure \ref{Wl}. In the limit of large $T$, the expectation value of $W_{\3Q}$ can
be written in the form

\begin{equation}\label{bloop}
\langle W_{\3Q}\rangle=\sum_{n=0}^\infty w_n\ep^{-V_{\3Q}^{(n)}T}
\,.
\end{equation}
Here $V_{\3Q}^{(0)}$ is the ground state energy, also called the three-quark potential, and the remaining $V_{\3Q}^{(i)}$'s are excited state energies. These are called hybrid three-quark potentials. 

The crucial fact underlying the lattice calculations is that these were performed in $SU(3)$ pure gauge theory. Because of this, the three-quark potential increases as quarks are pulled apart. By contrast, it is expected that in the presence of light quarks the potential flattens out. In string models of hadrons such a phenomenon is interpreted as string breaking \cite{stringP}. It is worth noting that in the $Q\bar Q$ system string breaking was established by numerical simulations \cite{bali}. In particular, the estimates of a scale (string breaking distance) characterizing this phenomenon were recently obtained for different light quark masses \cite{bulava}. 

The strong decay of a heavy meson into a pair of heavy-light mesons

\begin{equation}\label{mmode}
	Q\bar Q\rightarrow Q\bar q+\bar Q q
	\,
\end{equation}
can be interpreted as breaking a single string by light quark-antiquark pair creation. In the case of light quarks with equal masses, one way to describe the ground state energy of the $Q\bar Q$ system is to consider a two-state system with a model Hamiltonian \cite{drum}

\begin{equation}\label{HD}
{\cal H}=
\begin{pmatrix}
E_{\QQb} & g \\
g& 2E_{\Qqb} \\
\end{pmatrix}
\,.
\end{equation}
Here $E_{\QQb}$ is the energy of two separated heavy quark sources which are connected by a string and $2E_{\Qqb}$ is the energy of a non-interacting pair of heavy-light mesons. The off-diagonal element $g$ describes the mixing between these two states. The eigenvalues of this model Hamiltonian give the energy levels of the $Q\bar Q$ system. In this case a characteristic scale, called the string breaking distance, is naturally defined by \cite{drum,bulava}

\begin{equation}\label{lc-mes}
E_{\QQb}(\boldsymbol{\ell}_{\QQb})=2E_{\Qqb}
\,.
\end{equation}

In the string models of hadrons one thinks of a baryon as a system of three quarks connected by a Y-shaped string configuration \cite{stringP}. Such a picture implies the existence of several decay channels 

\begin{equation}\label{decay}
\begin{split}
&\,QQq+Q\bar q\\
\nearrow & \downarrow \\
\,QQQ\,\,\rightarrow & \,\,Qqq+2Q\bar q\,\\
\searrow & \downarrow\,\\
&\,\,\,qqq+3Q\bar q\
\,,
\end{split}
\!\!\!\xdownarrow{0.7cm}
\end{equation}
defined by a number of broken strings or, equivalently, by a number of light quark-antiquark pairs. To formalize the analogy with the mesonic case, one could consider a model Hamiltonian  

\begin{equation}\label{Hb}
{\cal H}=
\begin{pmatrix}
\,E_{\3Q} & g_1 & g_2 & g_3\\
g_1& E_{\QQq}+E_{\Qqb} &g_{12} & g_{13}\\
g_2 & g_{12} & E_{\Qqq}+2E_{\Qqb} & g_{23}\\
g_3 & g_{13} & g_{23} & E_{3q}+3	E_{\Qqb}\,\,\,
\end{pmatrix}
\end{equation}
for a four-state system. The $E$'s represent energies of quark sources connected by strings. Here we assume that hadrons are non-interacting, hence the total energies are the sums of the energies of the individual hadrons. The off-diagonal elements describe the mixing between these four states. 

The two important differences from the $Q\bar Q$ system are (i) that it is not clear how to define the string breaking distance by equating the diagonal elements of the model Hamiltonian. Indeed, the relative position of the heavy quarks, placed at vertices of a triangle, is defined by three parameters, and the equations can only say how the minimal energy state evolves in the parameter space. Transitions between such different minimal energy states occur at critical parameter values. In general, given the critical values, one can not determine the critical string distance without referring to any particular string model. Though the exceptions exist for some symmetric cases, as we will see in Secs.III and V; (ii) by equating the diagonal elements one can get at least five scales. If so, it is not a priori that the ground state energy depends only on one of them. 

The basic task of this paper is to further advance the use of effective string models in QCD. Here we continue our study of the static three-quark potentials \cite{a-3q}, elaborating on the phenomenon of string breaking. So far this issue has not been discussed in the literature, neither in the context of lattice gauge theory nor in the context of the gauge/string duality, particularly in AdS/CFT. The rest of the paper is organized as follows. For orientation, we begin in Sec.II by setting the framework and recalling some preliminary results. Then, in Sec.III, we look at two special geometric configurations of heavy quarks. An important feature of those is symmetry which allows one to actually determine the string breaking distance from the critical values of parameters without referring to any specific string model. We continue in Sec.IV with two type of geometries. These interpolate between a diquark limit, in which two heavy quarks are very close and the remaining one is far away from these two, and the geometries of Sec.III. In Sec.V we provide more details about the special geometries of Sec.III. Finally, we make concluding remarks in Sec.VI. Appendix A contains our notation and definitions. To make the paper more self-contained, we include many necessary results and technical details in Appendices B-D. 

\section{The model}
\renewcommand{\theequation}{2.\arabic{equation}}
\setcounter{equation}{0}
 
In the gauge/string duality, the expression for the expectation value of a baryonic Wilson loop can be put into a semiclassical form 

\begin{equation}\label{wilson}
\langle\,W_{\3Q}\,\rangle=\sum_n w_n\ep^{-S_n}
\,,
\end{equation}
 where $S_n$, whose relative weight is $w_n$, is expressed in terms of an energy of the string configuration $E_n$ and a time interval $T$ by $S_n=E_n T$. The $E_n$'s are the diagonal elements of the model Hamiltonian \eqref{Hb}. That is the key point which already allowed us to compare the results obtained for the $Q\bar Q$ system in \cite{a-strb} to those of lattice gauge theory.
 
For three colors, string configurations are built in analogy with tree diagrams of $\varphi^3$ field theory \cite{stringP}. The dictionary is as follows. A propagator now means a string which is represented by a solid line. A vertex means a string junction, nowadays called the baryon vertex. The baryon vertex always has three strings attached to it. Strings also end on quarks which play a role of sources. An important ingredient in describing string configurations on $\text{AdS}_5$-like geometries is a gravitational force. This force acts on all objects. In particular, it bends strings which are now represented as curves. Heavy quark sources are set on the boundary of five-dimensional space and light quark sources in its interior. 

To further illustrate these ideas, we employ a particular effective string model. Though it is not exactly dual to QCD, our reasons for pursuing this model are: (i) Because it would be good to gain some insight into any problems for which there are no predictions from phenomenology and the lattice, but which can be solved with the effective string model already at our disposal. (ii) Because the results provided by this model on the quark-antiquark and three-quark potentials are consistent with the lattice calculations and QCD phenomenology \cite{a-3q,az1}. (iii) Because analytic formulas are obtained by solving the model. (iv) Because the aim of our work is to make predictions which may then be tested by means of other non-perturbative methods, e.g., numerical simulations.

We will consider only the simplest class of 5-dimensional geometries which is an extension of the geometry used for successful modeling the heavy quark potentials in pure gauge theory. The extension is due to a scalar field which is responsible for light quarks at string endpoints and, as a consequence, for string breaking \cite{a-strb}.\footnote{This is an attempt to describe light sea quarks along the lines of \cite{son}, but in the framework of the worldsheet formalism. Since such a scalar signals an instability of a fundamental string, it seems natural to call it a tachyon. If so, it could correspond to an open string tachyon because of its action is a boundary term. One difficulty with this interpretation is that the scalar field has a real mass \cite{son}. So, this tachyon is not the usual tachyon around the string perturbative vacuum in flat space. We introduce a single field, because in what follows we will be interested only in the case of two light quarks of equal mass.} This background is of the form

\begin{equation}\label{metric}
ds^2=\ep^{\s r^2}\frac{R^2}{r^2}\Bigl(dt^2+d\vec x^2+dr^2\Bigr)
\,,
\qquad
{\text T}={\text T}(r)
\,,
\end{equation}
where $\text{T}$ is the scalar field. In the absence of it, the above metric reduces to a one-parameter deformation, with a deformation parameter $\s$, of that for the Euclidean $\text{AdS}_5$ space of radius $R$. The geometry has two important features:(i) aside from the boundary at $r=0$, there is a soft wall at $z=1/\sqrt{\s}$ which prevents strings from going deep into the bulk and (ii) a corresponding gravitational force has only a radial component. 

As noted above, we need three basic ingredients. The first is the fundamental string governed by the Nambu-Goto action 

\begin{equation}\label{NG}
S_{\text{\tiny NG}}=\frac{1}{2\pi\alpha'}\int d^2\xi\,\sqrt{\gamma^{(2)}}
\,.
\end{equation}
Here $\gamma$ is an induced metric, $\alpha'$ is a string parameter, and $\xi^i$ are worldsheet coordinates. 

The second ingredient is the baryon vertex. In the context of AdS/CFT correspondence it is a five brane \cite{witten}. At leading order in $\alpha'$ its dynamics is described by the action $S_{\text{vert}}={\cal T}_5\int d^6\xi\sqrt{\gamma^{(6)}}$, with ${\cal T}_5$ a brane tension and $\xi^i$  world-volume coordinates. Since the brane is wrapped on a five-dimensional internal space $\mathbf{X}$, the vertex looks point-like in Euclidean space \eqref{metric}. Following \cite{a-3q}, we pick a static gauge $\xi^0=t$ and $\xi^a=\theta^a$, with $\theta^a$ coordinates on $\mathbf{X}$. The action is then 

\begin{equation}\label{baryon-v}
S_{\text{vert}}=\tau_v\int dt \,\frac{\ep^{-2\s r^2}}{r}
\,,
\end{equation}
where $\tau_v$ is a dimensionless parameter defined by $\tau_v={\cal T}_5R\,\text{vol}(\mathbf{X})$ and $\text{vol}(\mathbf{X})$ is a volume of $\mathbf{X}$. 

The last ingredient which takes account of light quarks at string endpoints is the scalar field. Since we wish to mimic the $u$ and $d$ quarks of QCD which have equal masses, we add to the worldsheet action a boundary term $S_{\text{q}}=\int d\tau e\,\text{T}$. Such a term is usual for strings propagating in an open string tachyon background. The integral is over a worldsheet boundary parameterized by $\tau$ and $e$ is a boundary metric. As in \cite{a-strb}, we consider only the case of a constant background $\text{T}_0$ and worldsheets whose boundaries are lines in the $t$ direction. In that case, the action written in the static gauge is    

\begin{equation}\label{Sq}
S_{\text q}=\m\int dt \frac{\ep^{\frac{\s}{2}r^2}}{r}
\,,
\end{equation}
where $\m=R{\text T}_0$. This is obviously the action of a particle of mass ${\text T}_0$ at rest.

Finally, let us note that in practice it is not so convenient to use the constant factors in the above actions as the model parameters. Instead, we choose the model parameters as follows: $\g=\frac{R^2}{2\pi\alpha'}$, $\k=\frac{\tau_v}{3\g}$, and $\n=\frac{\m}{\g}$. 

\section{A first look at two special cases}
\renewcommand{\theequation}{3.\arabic{equation}}
\setcounter{equation}{0}

To gain an initial intuition about string breaking in the three-quark system, we consider two simple geometries. The first is a geometry with the heavy quarks are at the vertices of an equilateral triangle and the second with the heavy quarks are on a straight line with equal separation. The basic reason for considering these particular geometries is that the string breaking distance can be naturally defined in terms of the quark positions. In this section, we restrict to the decay mode $QQQ\rightarrow QQq+Q\bar q$. We return to this question in Sec.V, after describing the string configuration for $qqq$.
\subsection{Equilateral triangle geometry}

We begin with the equilateral triangle geometry. As usual, we place the heavy quarks at the boundary points of the five-dimensional space and consider static string configurations in which each quark is a string endpoint. Strings join at baryon vertices, or are terminated by light quarks in the interior.

 \subsubsection{String configurations}

\begin{figure}[htbp]
\centering
\includegraphics[width=5.9cm]{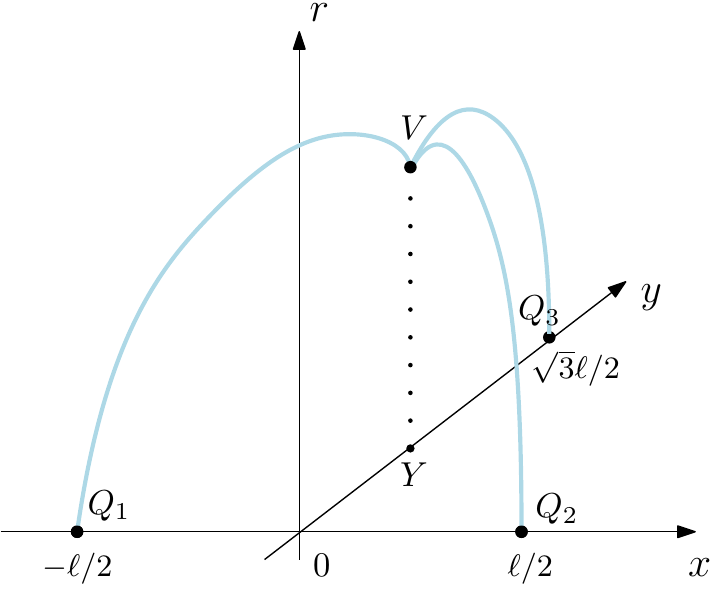}
\hspace{2cm}
\includegraphics[width=6.4cm]{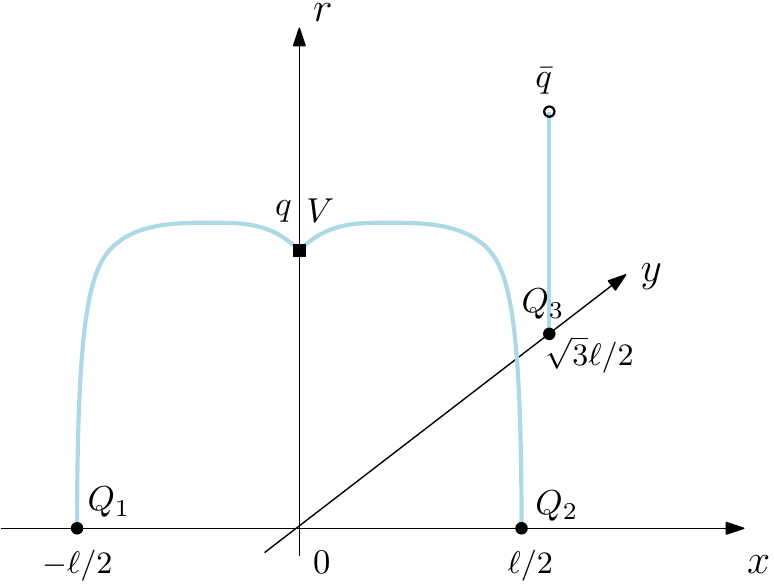}
\caption{{\small String configurations for the $QQQ$ system with a number of light quarks less or equal to $2$. The heavy quarks are at the vertices of the equilateral triangle of side $\ell$. Left: A connected configuration. Right: A disconnected configuration.}}
\label{etg}
\end{figure}

Since we are interested in a decay mode which is due to one pair of light quarks, we need to consider only two string configurations. These are shown in Figure \ref{etg}.\footnote{The configuration shown in the right panel is appropriate only for large $\ell$. A full set of the configurations for the $QQq$ system is as shown in Figure \ref{QQq}. For notation, see Appendix A.}  

First, let us consider the connected configuration. For the background geometry \eqref{metric}, the energy as a function of quark separation, can be written in parametric form \cite{a-3q}

\begin{equation}\label{E3Q-et}
\ell=\sqrt{\frac{3}{\s}}\,{\cal L}^-(\lambda,v)
\,,\qquad
E_{\3Q}=3\g\sqrt{\s}
\biggl(
{\cal E}^-(\lambda,v)
+
\k\frac{\ep^{-2v}}{\sqrt{v}}
\biggr)
+3c
\,,
\end{equation}
where $v$ is a parameter and $c$ is a normalization constant (see Appendix B). The functions ${\cal L}^-$, ${\cal E}^-$ are defined in Appendix A. $\lambda$ is a function of $v$ such that

\begin{equation}\label{lambda-etg}
\lambda(v)=-\text{ProductLog}\Bigl[-v\ep^{-v}\bigl(1-\k^2(1+4v)^2\ep^{-6v}\bigr)^{-\frac{1}{2}}\Bigr]
\,.
\end{equation}
Here $\text{ProductLog}$ denotes the ProductLog function \cite{wolf}. The parameter $v$  runs from $0$ to $\vet$, where $\vet$ is a solution to $\lambda(v)=1$ (see also Eq.\eqref{vast} below). 

For future reference, note that the large $\ell$ behavior of $E_{\3Q}$ is 

\begin{equation}\label{E3Q-etg-large}
E_{\3Q}=\sqrt{3}\sigma\ell-3\g\sqrt{\s}I_{\3Q}+3c+o(1)
\,,
\end{equation}
with
\begin{equation}\label{sigma}
I_{\3Q}={\cal I}(\vet)-\k\frac{\ep^{-2\vet }}{\sqrt{\vet}}
\,,\qquad
\sigma=\ep\g\s
\,.
\end{equation}
Here $\sigma$ is the string tension and ${\cal I}$ is defined by Eq.\eqref{I}. As one can easily recognize, such a behavior is governed by the $Y$-law.\footnote{In the context of lattice gauge theory the $Y$-law was discussed in \cite{bali,3Q-lattice}.}
 
To get further, consider the disconnected configuration. In the gauge/string duality it is interpreted as a non-interacting pair of hadrons. So, the energy of the configuration is the sum of $E_{\QQq}$ and $E_{\Qqb}$, whose explicit expressions are given in Appendix D. For our purposes, we will need to know about its large $\ell$ behavior. Using the formulas \eqref{EQqb} and \eqref{EQQq-large}, we get 

\begin{equation}\label{Edis-large}
E_{\QQq}+E_{\Qqb}=\sigma\ell+
\g\sqrt{\s}\Bigl({\cal Q}(q)+\n\frac{\ep^{\oh q}}{\sqrt{q}}-2I_{\QQq}\Bigr)
+3c+o(1)
	\,.
	\end{equation}
Here the function ${\cal Q}$ is defined by \eqref{Q}, $q$ is a solution to equation \eqref{q}, and $I_{\QQq}$ is given by \eqref{IQQq}. Thus, we see again that the asymptotic behavior for large $\ell$ is described by the $Y$-law.

\subsubsection{String breaking}

\begin{figure}[H]
\centering
\includegraphics[width=7.35cm]{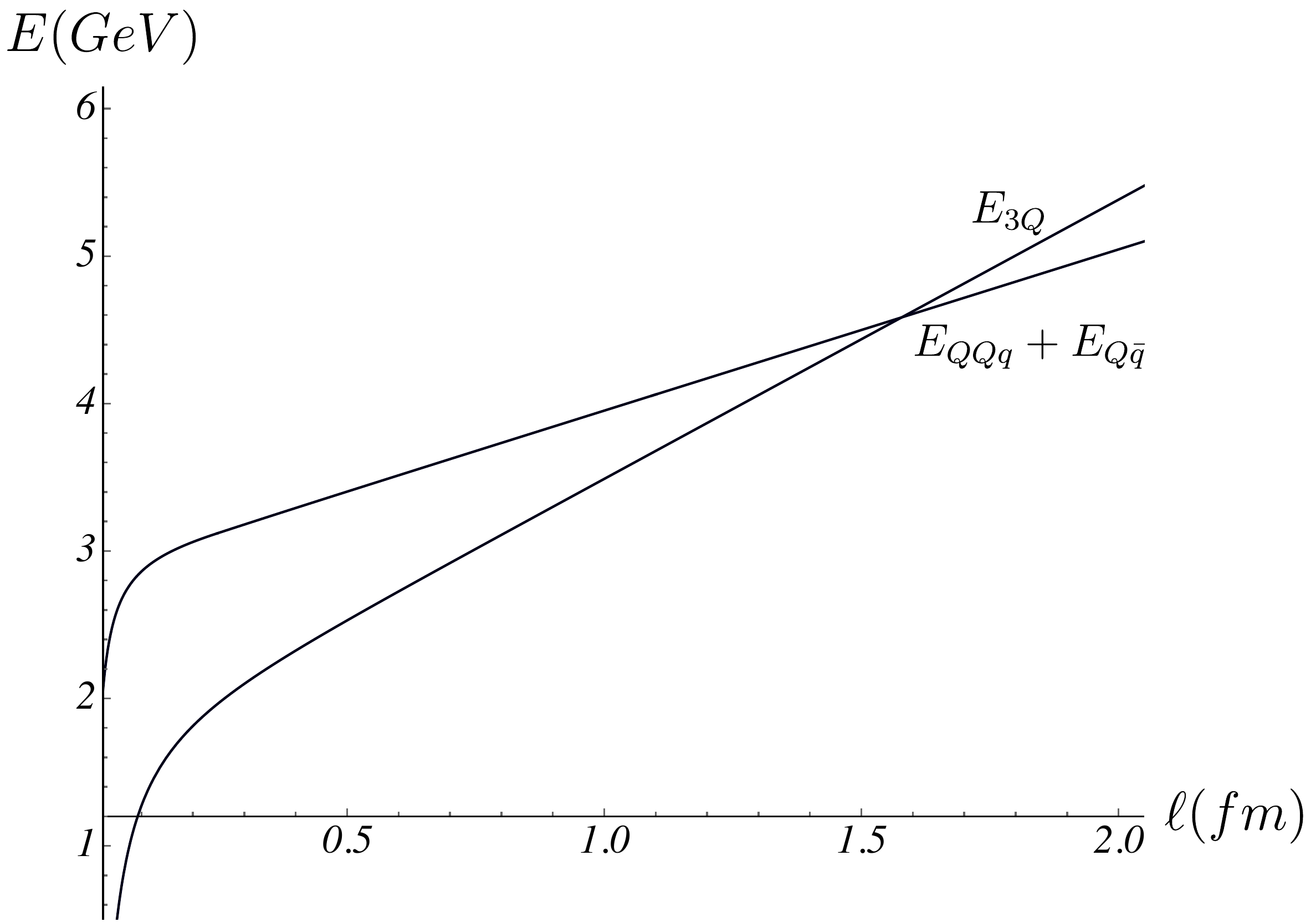}
\caption{{\small $E_{\3Q}$ and $E_{\QQq}+E_{\Qqb}$ vs $\ell$ in the case of the equilateral triangle geometry for $\k=-0.102$ and $c=0.623\,\text{GeV}$.}}
\label{EE-etg}
\end{figure}

Given the expressions for the energies, we can find a characteristic scale for this decay mode by solving the equation $E_{\3Q}(\ell)=E_{\QQq}(\ell)+E_{\Qqb}$. In general this is a complicated problem, best done numerically, but what saves the day is that $\ell_c$ is an infrared scale. If so, then the equation is linear in very good approximation. To see that this is the case, we plot the energies in Figure \ref{EE-etg}. We use the parameter set $L$ defined below. Indeed, both functions behave de facto 

 \noindent as linear functions for $\ell\gtrsim 0.5\,\text{fm}$. Given this, the critical value of $\ell$ is\footnote{Here and below, the superscript $(1)$ indicates that the scale is set by equating $E_{\3Q}$ to $E_{\QQq}+E_{\Qqb}$, where $1$ stands for one pair of $q\bar q$.} 

\begin{equation}\label{lc}
\ell_c^{(1)}=\frac{1+\sqrt{3}}{2\ep\sqrt{\s}}
\Bigl(
{\cal Q}(q) +\n\frac{\ep^{\oh q}}{\sqrt{q}}+3I_{\3Q}-2I_{\QQq}
\Bigr)
\,,
\end{equation}
as it follows from the $Y$-laws \eqref{E3Q-etg-large} and \eqref{Edis-large}. It is finite and independent of $c$, as expected.

There is a nontrivial issue concerning the question how to define the notion of a string length in five dimensional models which is consistent with QCD in four dimensions. In four dimensions, the string description is valid at large quark separations, and as a result, the energy of the $QQQ$ system is described by the $Y$-law. In this case, the string length is naturally defined as a distance between its endpoints, one of which is attached to a heavy quark and the other to a string junction. All the strings in question are straight and the junction is at the Fermat point of a triangle $Q_1Q_2Q_3$. The situation becomes less transparent when dealing with five dimensional string models motivated by the gauge/string duality. The problem here is that a baryon vertex is in the interior, while heavy quarks are on the boundary of a five-dimensional curved space. To make a contact with QCD, it is natural to define the string length as a distance between a string endpoint attached to a heavy quark and a projection of the baryon vertex onto the boundary, i.e. by $\vert Q_i Y\vert$ (see Figure \ref{etg}). The good reason for this is that in those models the $Y$-law is written precisely in terms of such defined string length \cite{a-3q0}. 

In the present case, the definition of the string length is dictated by symmetry. The point $Y$ is always at the triangle center which coincides with its Fermat point, and hence $\vert Q_iY\vert=\ell/\sqrt{3}$. Thus, $\boldsymbol{\ell}_{\3Q}^{(1)}=\vert Q_iY\vert_c=\ell_c^{(1)}/\sqrt{3}$. Explicitly,

\begin{equation}\label{Lc-etg}
\boldsymbol{\ell}_{\3Q}^{(1)}=\frac{3+\sqrt{3}}{6\ep\sqrt{\s}}
\Bigl(
{\cal Q}(q) +\n\frac{\ep^{\oh q}}{\sqrt{q}}+3I_{\3Q}-2I_{\QQq}
\Bigr)
\,.
\end{equation}
This is the formula we will need to make some preliminary estimates of the string breaking distance in subsection C. 

\subsection{Symmetric collinear geometry}

Our second special example is that of the collinear geometry. Without loss of generality, we may place heavy quarks on the $x$-axis at points $(-\ell,0,\ell)$. As before, these quarks are string endpoints. The other endpoints are terminated at light quarks and vertices in the interior to form color singlets.

\subsubsection{String configurations}

\begin{figure}[htbp]
\centering
\includegraphics[width=6.5cm]{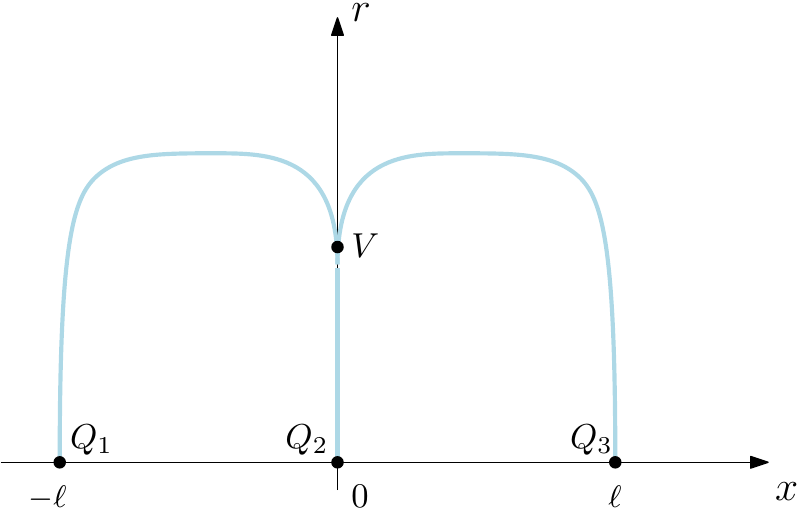}
\caption{{\small A connected collinear configuration. It is symmetric under reflection through the $r$-axis. So, $Q_2$ is the projection of $V$ onto the $x$-axis.}}
\label{col-c}
\end{figure}

We begin by considering the configuration of Figure \ref{col-c}. It has been previously 
described in \cite{a-3q}, where the explicit formula for the energy was derived. When written in terms of the functions defined in Appendix A, this formula is 

\begin{equation}\label{E3Q-col}
\ell=\frac{1}{\sqrt{\s}}\,{\cal L}^-(\lambda,v)
\,,\qquad
E_{\3Q}=\g\sqrt{\s}\Bigl(
2{\cal E}^-(\lambda,v)
+
3\k\frac{\ep^{-2v}}{\sqrt{v}}
+
{\cal Q}(v)
\Bigr)+3c
\,.
\end{equation}
Here $\lambda$ is a function of $v$ which is given by  

\begin{equation}\label{lambda-col}
\lambda(v)=-\text{ProductLog}
\biggl(-\frac{2}{\sqrt{3}}v\,\ep^{-v}
\Bigl(1+2\k(1+4v)\ep^{-3v}-3\k^2(1+4v)^2\ep^{-6v}\Bigr)^{-\tfrac{1}{2}}
\biggr)
\,.
\end{equation}
The parameter $v$ is varying from $0$ to $\bar v$, where $\bar v$ is a solution to 

\begin{equation}\label{barv}
\frac{4}{3}v^2\ep^{2(1-v)}=1+2\k(1+4v)\ep^{-3v}-3\k^2(1+4v)^2\ep^{-6v}
	\,
\end{equation}
or, equivalently, to $\lambda(v)=1$. 

Just as in the above example, the large $\ell$ behavior is governed by the $Y$-law which in the present case takes the form

\begin{equation}\label{E3Q-col-large}
E_{\3Q}(\ell)=2\sigma\ell-2\g\sqrt{\s}\bar I_{\3Q}+3c+o(1)\,,
\quad\text{with}\quad
\bar I_{\3Q}=
\mathcal{I}(\bar v)
-
\oh\mathcal{Q}(\bar v)
-
\frac{3}{2}\k\frac{\ep^{-2\bar v}}{\sqrt{\bar v}}
\,.
\end{equation}
Note that for this geometry the Fermat point is coincident with $Q_2$. 

\begin{figure}[htbp]
\centering
\includegraphics[width=7cm]{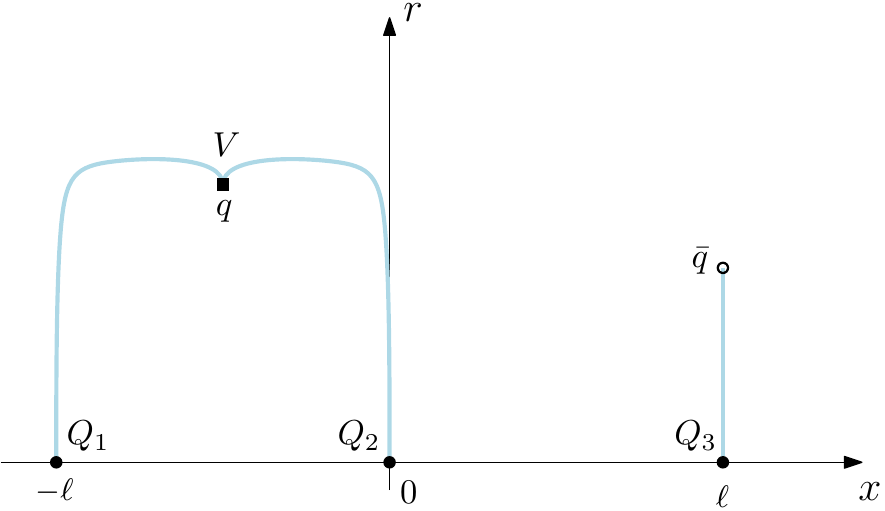}
\hspace{1.5cm}
\includegraphics[width=6.5cm]{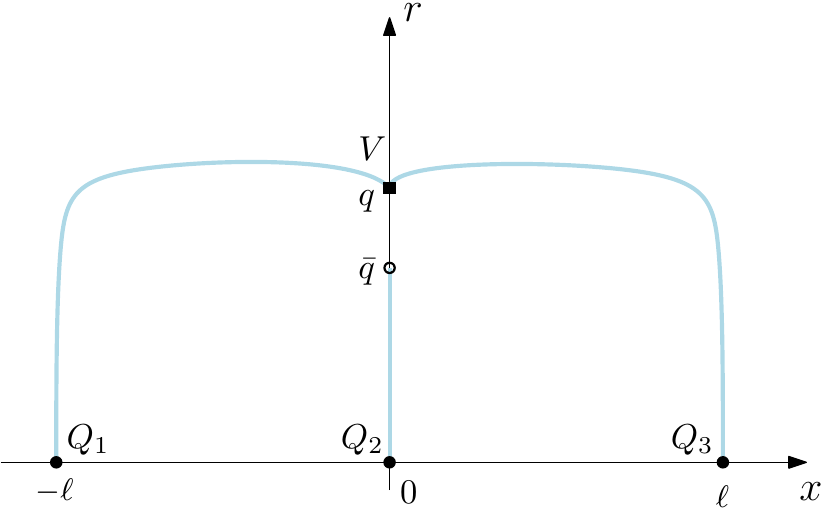}
\caption{{\small Typical disconnected collinear configurations. In the configurations shown here, $\ell$ is large enough.}}
\label{col-d}
\end{figure}

Now let us consider the disconnected configurations shown in Figure \ref{col-d}. Here, unlike the first example, there are      
two possible configurations. The configuration shown in the left panel corresponds to the case when one of the side strings breaks down, whereas the configuration shown in the right panel to the case when the middle string breaks down. Among of them, the first provides a dominant contribution to the baryonic Wilson loop as its energy is smaller. The simple reason for this is that its $QQq$ subsystem has a smaller length, $\ell$ rather than $2\ell$. Since we assume that hadrons do not interact with each other, the dominant configuration has the same energy as the disconnected configuration shown in Figure \ref{etg}. Hence its large $\ell$ behavior is as given by Eq.\eqref{Edis-large}.  

\subsubsection{String breaking}

\begin{figure}[htbp]
\centering
\includegraphics[width=7cm]{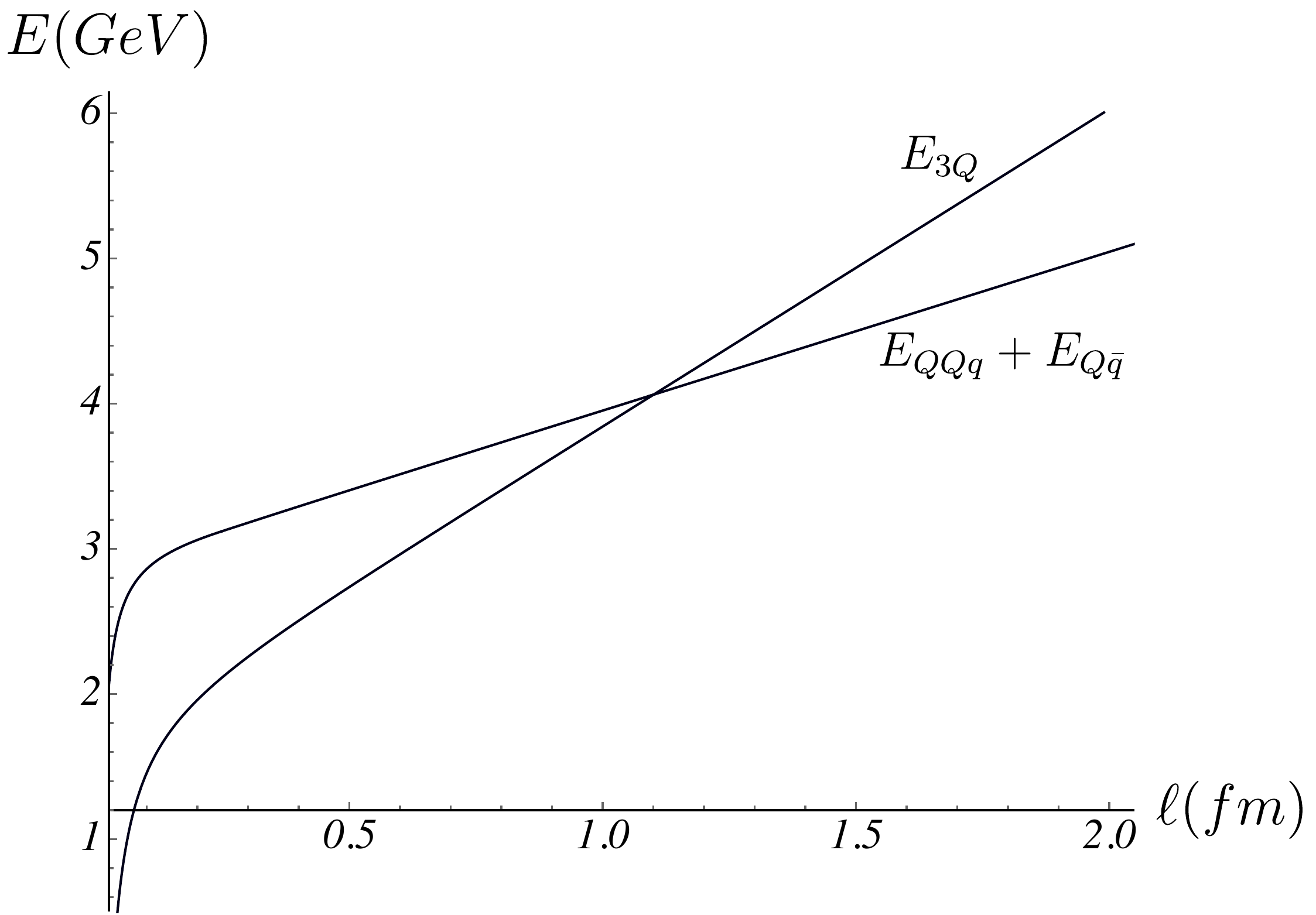}
\caption{{\small $E_{\3Q}$ and $E_{\QQq}+E_{\Qqb}$ vs $\ell$ for the symmetric collinear geometry, plotted here for the parameter set $L$ with $\k=-0.102$ and $c=0.623\,\text{GeV}$.}}
\label{EE-col}
\end{figure}

We are now in position to estimate the string breaking distance for this geometry. As before, we do so by solving the equation $E_{\3Q}(\ell)=E_{\QQq}(\ell)+E_{\Qqb}$. This problem is  simple because the linear approximation is applicable for large $\ell$, as seen from the plot in Figure \ref{EE-col}. Given a critical value $\ell_c^{(1)}$, the string breaking distance is $\boldsymbol{\ell}_{\3Q}^{(1)}=\ell_c^{(1)}$ as dictated  

\noindent by symmetry. A short calculation, using the formulas \eqref{E3Q-col-large} and \eqref{EQQq-large}, shows that 

\begin{equation}\label{lc1-col}
\boldsymbol{\ell}_{\3Q}^{(1)}=\frac{1}{\ep\sqrt{\s}}
\Bigl(
{\cal Q}(q) +\n\frac{\ep^{\oh q}}{\sqrt{q}}+2\bar I_{\3Q}-2I_{\QQq}
\Bigr)
\,,
\end{equation}
which is the analog of Eq.\eqref{Lc-etg}.

\subsection{Some estimates}

Let us make some estimates. In doing so, we use the two parameter sets suggested in \cite{a-strb}. The first is mainly a result of fitting the lattice QCD data to the string model we are considering. So, we call it $L$ for brevity. The value of $\s$ is fixed from the slope of the Regge trajectory of $\rho(n)$ mesons in the soft wall model with the geometry \eqref{metric}. This gives $\s=0.450\,\text{GeV}^2$ \cite{a-q2}. Then, using \eqref{sigma}, we obtain $\g=0.176$ by fitting the value of the string tension $\sigma$ to its value in \cite{bulava}. The parameter $\n$ is adjusted to reproduce the lattice result for the string breaking distance in the $Q\bar Q$ system. With $\boldsymbol{\ell}_{\QQb}=1.22\,\text{fm}$ \cite{bulava}, this gives $\n=3.057$. In \cite{a-3q}, the value of $\k$ is adjusted to fit the three-quark potential to the lattice data for pure $SU(3)$ gauge theory. So far there is no such data available for QCD with two dynamical quarks. We will use two different parameter values: $\k=-0.102$ and $\k=-\frac{1}{4}\ep^{\frac{1}{4}}$. These parameter values are of special interest. The first comes from matching to the Lipkin rule $E_{\QQ}=\oh E_{\QQb}$ \cite{lipkin}, but at the same time it is rather close to the value of $-0.083$ used in \cite{a-3q} to describe the lattice data for the static three-quark potentials in $SU(3)$ pure gauge theory \cite{3Q-lattice}. The reasons for the second value are due to the particular model we are using here.\footnote{We discuss them in Sec.V.}

In \cite{bulava}, the calculations were done at unphysical pion mass $m_{\pi}=280\,\text{MeV}$. Because of this and in view of possible applications to phenomenology, we now consider the second parameter set denoted by $P$. In that case, the values of $\s$ and $\g$ are extracted from the quarkonium spectrum obtained by using the heavy quark potential derived from the model we are considering \cite{az1}. This is self-consistent, and gives $\s=0.15\,\text{GeV}^2$ and $\g=0.44$ \cite{PC}.\footnote{At first glance, the difference between these parameter sets is significant. However, what is relevant for making the estimates of string breaking distances is the value of the string tension \eqref{sigma}. With the $L$ set, it is about $0.215\,\text{GeV}^2$, while with $P$ about $0.285 \,\text{GeV}^2$. So, the values are different but not greatly so.} Then we determine $\n$ from the condition $E_{\Qqq}-E_{\Qqb}=M_{\Lambda_c^+}-M_{D^0}\approx 420\,\text{MeV}$. It results in $\n=1.589$. We use $\k=-0.102$ and $\k=-\tfrac{1}{4}\ep^{\frac{1}{4}}$, as before.  

Having set the parameters, one can use Eqs.\eqref{Lc-etg} and \eqref{lc1-col} to estimate the string breaking distance $\boldsymbol{\ell}_{\3Q}^{(1)}$. The results are summarized in Table \ref{estimates}.

\begin{table*}[htb]
\renewcommand{\arraystretch}{2}
\centering 	\scriptsize
\begin{tabular}{lccccccr}				
\hline
Geometry        ~&~  $\s\,(\text{GeV}) $ ~&~~~ $\g$  ~~~&~~~~ $\n$ ~~~~&~~~~ $\k$ ~~~&~ $\boldsymbol{\ell}_{\3Q}^{(1)}\,(\text{fm})$  ~&~ $\boldsymbol{\ell}_{\QQb}\,(\text{fm})$ ~&~ $\frac{\boldsymbol{\ell}_{\3Q}^{(1)}}{\boldsymbol{\ell}_{\QQb}}$
\rule[-3mm]{0mm}{8mm}
\\
\hline \hline
{\tt triangle} & 0.450  & 0.176 & 3.057 & -0.102 & 0.910 & 1.220 & 0.746 \\
{\tt triangle} & 0.450 & 0.176 & 3.057& -$\frac{1}{4}\ep^{\frac{1}{4}}$ & 0.911 & 1.220 & 0.747 \\
{\tt triangle}  & 0.150 & 0.440 & 1.589 & -0.102 & 0.810 & 1.074 & 0.754 \\
{\tt triangle}  & 0.150 & 0.440 & 1.589 & -$\frac{1}{4}\ep^{\frac{1}{4}}$ &0.810 & 1.074 & 0.754 \\  
\hline
{\tt collinear} & 0.450  & 0.176 & 3.057 & -0.102 & 1.100 & 1.220 & 0.902 \\
{\tt collinear} & 0.450 & 0.176 & 3.057& -$\frac{1}{4}\ep^{\frac{1}{4}}$ & 1.180 & 1.220 & 0.967 \\
{\tt collinear}  & 0.150 & 0.440 & 1.589 & -0.102 & 0.934 & 1.074 & 0.870 \\
{\tt collinear}  & 0.150 & 0.440 & 1.589 & -$\frac{1}{4}\ep^{\frac{1}{4}}$ &1.070 & 1.074 & 0.996 \\  
\hline \hline
\end{tabular}
\caption{ \small Results for the string breaking distance $\boldsymbol{\ell}_{\3Q}^{(1)}$ obtained from the $L$ and $P$ parameter sets at two values of $\k$.}
\label{estimates}
\end{table*}

\newpage 
We conclude this subsection by making a few remarks. 

\noindent (i) We have used the different parameter sets to assess the robustness of the conclusion that, in contrast to the string tension, the string breaking distance is not universal. In the $QQQ$ system it may be smaller than that in the $Q\bar Q$ system and, moreover, it may depend on geometry, i.e. on the positions of the three heavy quarks in space. 

\noindent (ii) We included the results of \cite{a-strb} for the string breaking distance in the $Q\bar Q$ system because it is of physical interest to compare both systems. The explicit formula reads 

\begin{equation}\label{Lc-mes}
	\boldsymbol{\ell}_{\QQb}=\frac{2}{\ep\sqrt{\s}}
\Bigl(
{\cal Q}(q)+\n\frac{\ep^{\oh q}}{\sqrt{q}}+I_0
\Bigr)
\,,
\end{equation}
where the constant $I_0$ is defined in Appendix A. Note that it allows one to get rid of the explicit dependence of $\s$ and $\g$ in the ratio $\frac{\boldsymbol{\ell}_{\3Q}^{(1)}}{\boldsymbol{\ell}_{\QQb}}$. So, in contrast to the triangle geometry, a stronger $\k$-dependence is seen for the collinear geometry. 

\noindent (iii) It is interesting to make an estimate of the quark mass in our model. In \cite{son}, it was assumed that the quark mass is simply related to the parameter ${\text T}_0$ in Eq.\eqref{Sq}, namely $m={\text T}_0$. However, ${\text T}_0$ does not enter explicitly in the formulas for the string breaking distance because it is combined with $R$ to form a dimensionless product. The unknown parameter $R$ makes a direct estimate impossible. Yet what saves the day in the case with the $L$ parameter set is that in \cite{bulava} the sum of the quark masses ($m_{u/d}+m_s$) is approximately equal to its physical value.\footnote{We thank F. Knechtli for pointing this out.} This allows us to estimate the value of $R$ and hence the quark masses. The details are as follows. First, we extend the model by adding the strange quark. Then, using \eqref{Lc-mes}, we get $\n_s=3.239$ by fitting the value of the string breaking distance associated with a pair of strange quarks to $1.29\,\text{fm}$ from \cite{bulava}. Finally from the expression $\n=\frac{R}{\g} m$ specialized to the $u/d$ and $s$ quarks, we immediately find $m_{u/d}=\frac{m_{u/d}+m_s}{\n+\n_s}\n$. With $m_{u/d}+m_s=96\,\text{MeV}$, this gives $m_{u/d}=46.6\,\text{MeV}$. What is an appropriate value of $m_{u/d}$ for the $P$ set? Suppose that the ratio $\frac{R}{\g}$ is universal. In that case, the mass will be $m_{u/d}=23.5\,\text{MeV}$. This is of course a crude estimate, but it is in agreement with the fact that the string breaking distance for the $P$ set is smaller than that for the $L$ set. One may wonder about the interpretation of $m$. The answer could be that it is a running quark mass in the IR regime at the energy scale of order $\Lambda_{\text{\tiny QCD}}$. The reason for this is that the worldsheet action \eqref{Sq} was introduced to describe the IR phenomenon of string breaking. Of course, further work is needed to clarify this issue.

\noindent (iv) This is not the whole story because of the other modes in \eqref{decay}. These modes are discussed in Sec.V, but with the same conclusion.

\section{More general geometries}
\renewcommand{\theequation}{4.\arabic{equation}}
\setcounter{equation}{0}

Now we will describe two types of geometries that allow us to see how $\boldsymbol{\ell}_{\3Q}^{(1)}$ decreases from the value $\boldsymbol{\ell}_{\QQb}$ to the values we found in the previous Section. The basic idea is to start from a special case in which two heavy quarks are very close and one is far away from these two, or in other words from the diquark limit. In this limit the equation $E_{\3Q}=E_{\QQq}+E_{\Qqb}$ can be easily solved. From heavy quark-diquark symmetry \cite{wise}, it follows that $E_{\3Q}=E_{\QQb}+E_{\QQ}$ and $E_{\QQq}=E_{\text{\tiny q}\bar{\text{\tiny Q}}}$+$E_{\QQ}$. After substituting those in the equation, we find $E_{\QQb}=E_{\Qqb}+E_{\text{\tiny q}\bar{\text{\tiny Q}}}$ whose solution is $\boldsymbol{\ell}_{\QQb}$. This implies that in the diquark limit $\boldsymbol{\ell}_{\3Q}=\boldsymbol{\ell}_{\QQb}$. Then we depart from this limit by more and more separating the neighbor quarks until we reach one of the geometries of Section III. Importantly, the decay mode $QQQ\rightarrow QQq+Q\bar q$ is dominant as long as the separation is not large enough. In other words, the two remaining strings are short and, therefore, do not break down. So, one can identify $\boldsymbol{\ell}_{\3Q}^{(1)}$ with the string breaking distance seen from the ground state energy because this scale is the only possible measure for string breaking.

\subsection{Isosceles triangle geometry}

\begin{figure}[htbp]
\centering
\includegraphics[width=5.25cm]{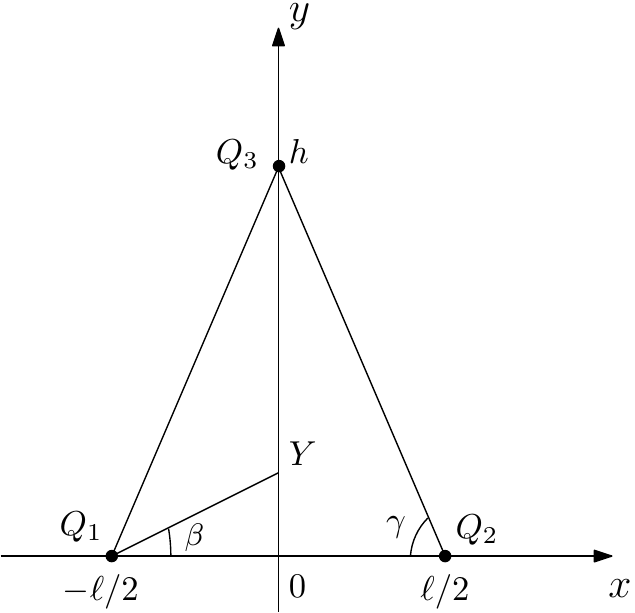}
\caption{{\small An isosceles triangle of base length $\ell$ and height $h$.}}
\label{ist}
\end{figure}

First consider a geometry in which the heavy quarks are at the vertices of an isosceles triangle. Without loss of generality, we can place the triangle in the $xy$-plane as shown in Figure \ref{ist}. The base angle $\gamma$ runs from $\frac{\pi}{3}$ to $\frac{\pi}{2}$. The 
former corresponds to the equilateral triangle geometry we discussed in the previous section, and the latter to the diquark limit, but with small enough $\ell$. For convenience, we list the formulas   

\begin{equation}\label{l-h}
\ell=2\cos\beta\,\vert Q_1Y\vert\,,
\qquad
	h=\sin\beta\vert Q_1Y\vert+\vert Q_3Y\vert
	\,,\qquad
\tan\gamma=\tan\beta+\frac{1}{\cos\beta}\frac{\vert Q_3Y\vert}{\vert Q_1Y\vert}
	\,,
\end{equation}
all of which are simple, but useful.

\subsubsection{Connected string configurations}

\begin{figure}[htbp]
\centering
\includegraphics[width=5.7cm]{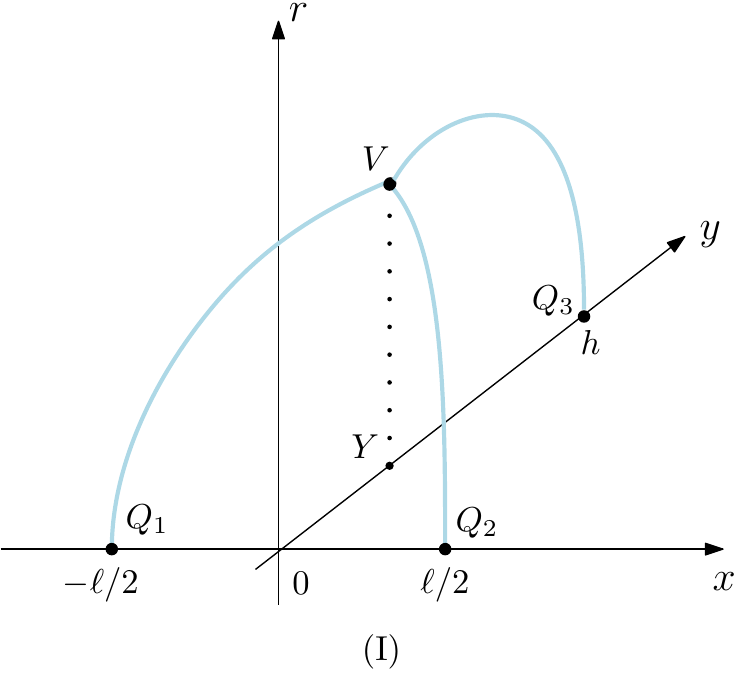}
\hspace{2cm}
\includegraphics[width=6.3cm]{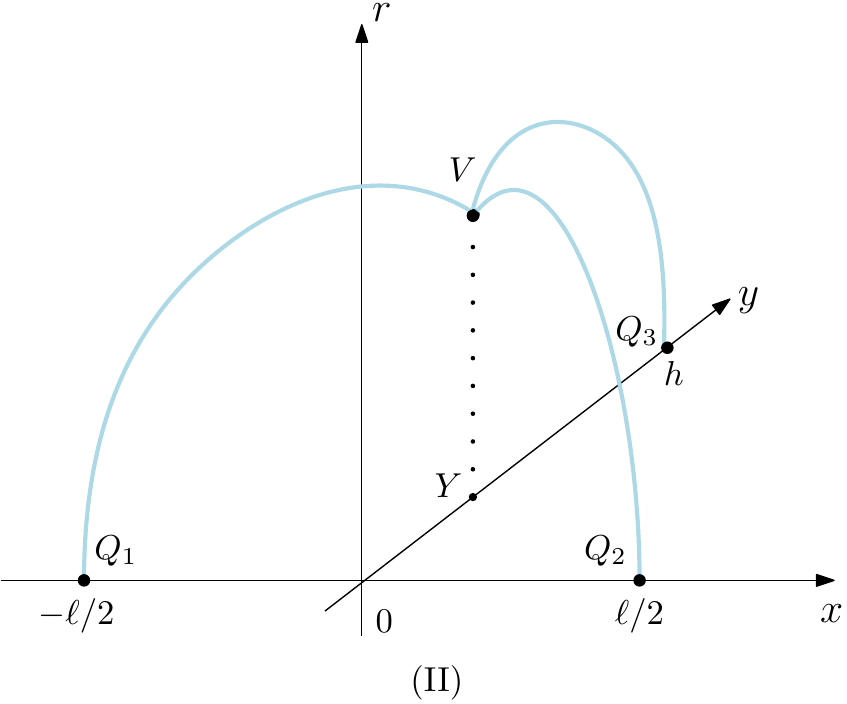}
\caption{{\small Connected string configurations in which $Q_3$ is away from $Q_1$ and $Q_2$. These are symmetric under reflection in the $yr$-plane. Left: The case $\alpha_1\geq0$. Right: The case $\alpha_1\leq0$. Here $\alpha_1$ is a tangent angle at $V$ for the string stretched between $Q_1$ and $V$.}}
\label{istg}
\end{figure}

We are interested in the situation when a string connected to $Q_3$ breaks down. For this to be the case, the string should be long enough. In practice, this means that the string is of type shown in Figure \ref{strings} on the right. If so, then there are only two possible string configurations. These are illustrated in Figure \ref{istg}.
Such configurations have not been studied in the literature, except for the diquark limit \cite{a-3q}. So, it is convenient to begin with this limiting case and then go further by increasing the base length $\ell$. If so, then the corresponding configuration is that shown in the Figure on the left. 

Given the configuration, we can use the general formulas \eqref{l+} and \eqref{l-} to write the distances between the points $Q_i$ and $Y$ as 

\begin{equation}\label{QiY+}
\vert Q_1Y\vert=\vert Q_2Y\vert=
\frac{1}{\sqrt{\s}}{\cal L}^+(\alpha_1,v)
\,,\qquad 
\vert Q_3Y\vert=\frac{1}{\sqrt{\s}}{\cal L}^-(\lambda_3,v)
\,.
\end{equation}
By symmetry, the shape of strings connected to $Q_1$ and $Q_2$ is the same. This implies that $\alpha_1=\alpha_2$. 

The string energies can be read off from the expressions \eqref{E+} and \eqref{E-}. Combining those with 
the gravitational energy of the vertex, we find the total energy of the configuration

\begin{equation}\label{EiY+}
E_{\3Q}= \g\sqrt{\s}
\Bigl(
2{\cal E}^+(\alpha_1,v)
+
{\cal E}^-(\lambda_3,v)
+
3\k\frac{\ep^{-2v}}{\sqrt{v}}\,
\Bigr)
\,
+
3c
\,.
\end{equation}

With the help of the relation \eqref{v-lambda} and gluing conditions \eqref{fbv}, the angles $\alpha_1$, $\alpha_3$, and $\beta$ are explicitly written as

\begin{equation}\label{angiY+}
\sin\alpha_1=\oh
\Bigl(
3\k(1+4v)\ep^{-3v}
+
\bigl[1-\cos^2\hspace{-1mm}\alpha_3\bigr]^{\oh}
\Bigr)
\,,
\qquad
\cos\alpha_3=\frac{v}{\lambda_3}\ep^{\lambda_3-v}
\,,\qquad
\sin\beta=\oh\cos\alpha_3\bigl[1-\sin^2\hspace{-1mm}\alpha_1\bigr]^{-\oh}
\,.	
\end{equation}
So the result for $E_{\3Q} (\ell,h)$ can be conveniently expressed in terms of two parameters $(v,\lambda_3)$. 

Now we wish to look at the behavior of $E_{\3Q}$ in the diquark limit: small $\ell$ and large $\vert Q_3Y\vert$. This case was discussed in relation to the static quark-quark potential in \cite{a-3q}. A starting point is that one has to analyze the parametric equations in a small region near the point $(0,1)$. We do not need all the results of the analysis. What is important for us to know is the behavior of $\ell$. It behaves as 

\begin{equation}\label{l-v}
\ell=\sqrt{\frac{v}{\s}}\bigl(\ell_0+o(1)\,\bigr)
\,.
\end{equation}
Here $l_0=\frac{1}{2}\xi^{-\frac{1}{2}}B\bigl(\xi^2;\tfrac{3}{4},\tfrac{1}{2}\bigr)$, $\xi=\frac{\sqrt{3}}{2}(1-2\k-3\k^2)^{\frac{1}{2}}$, and $B(z;a,b)$ is the incomplete beta function.

As we saw in Sec.III, to analyze string breaking, it helps to know that this phenomenon occurs in IR limit in which the linear approximation \eqref{E13} turns out to be quite accurate. Since we assume that the string ended on $Q_3$ breaks down, we should take the limit $\lambda_3\rightarrow 1$. In that limit, Eq.\eqref{EiY+} for the energy becomes 

\begin{equation}\label{EL+}
E_{\3Q}=\sigma \vert Q_3Y\vert-\g\sqrt{\s}
\Bigl(
{\cal I}(v)
-2{\cal E}^+(\alpha_1,v)
-3\k\frac{\ep^{-2v}}{\sqrt{v}}\,
\Bigr)
+
3c+o(1)\,,
\end{equation}
and similarly Eqs.\eqref{angiY+} for the auxiliary angles become 

\begin{equation}\label{anglesdi1}
\sin\alpha_1=\oh
\biggl(
3\k(1+4v)\ep^{-3v}
+
\Bigl[1-v^2\ep^{2(1-v)}\Bigr]^{\oh}
\biggr)
\,,\qquad
\cos\alpha_3=v\ep^{1-v}\,,
\qquad
\sin\beta=\oh v \ep^{1-v}
\Bigl[1-\sin^2\hspace{-1mm}\alpha_1 \Bigr]^{-\oh}
\,.	
\end{equation}
The parameter $v$ varies from $0$ to $\vit$, where $\vit$ is a solution to equation

\begin{equation}
	3\k(1+4v)\ep^{-3v}
+
\Bigl[1-v^2\ep^{2(1-v)}\Bigr]^{\oh}=0
\,.
\end{equation}
The solution $\vit$ has a clear meaning. The angle $\alpha_1$ precisely vanishes at $v=\vit$. As a result, for larger values of $v$ the string configuration becomes as shown in  Figure \ref{istg} on the right. What is, however, important for our purposes is that $\ell(v)$ is an increasing function of $v$. Thus the transition between these two types of configurations gives rise to larger values of $\ell$. 

Similarly, we may also analyze the remaining configuration. From \eqref{l-} it follows that 

\begin{equation}\label{QiY-}
\vert Q_1Y\vert=\vert Q_2Y\vert=\frac{1}{\sqrt{\s}}{\cal L}^-(\lambda_1,v)
\,,\qquad 
\vert Q_3Y\vert=\frac{1}{\sqrt{\s}}{\cal L}^-(\lambda_3,v)
\,.
\end{equation}
Because of the reflection symmetry, $\lambda_2$ is equal to $\lambda_1$. 

The expression for the energy is simply obtained by replacing ${\cal E}^+$ with ${\cal E}^-$ in \eqref{EiY+}. So, 

\begin{equation}\label{EiY-}
E_{\3Q}=\g\sqrt{\s}
\Bigl
(\sum_{i=1}^3
{\cal E}^-(\lambda_i,v)
+
3\k\frac{\ep^{-2v}}{\sqrt{v}}\,\Bigr)
+3c
\,.
\end{equation}
Combining the formulas \eqref{v-lambda} and \eqref{lambda} with the gluing conditions \eqref{fbv} gives 

\begin{equation}\label{lambdadi2}
\lambda_1=-\text{ProductLog}\biggl(-v\,\ep^{-v}
\biggl[1-\frac{1}{4}\biggl(3\k(1+4v)\ep^{-3v}+
\Bigl[1-\frac{v^2}{\lambda_3^2}\ep^{2(\lambda_3-v)}\Bigr]^{\oh}\biggr)^2\,\biggr]^{-\oh}
\,\biggr)
\,,\qquad
\sin\beta=\frac{\lambda_1}{2\lambda_3}\ep^{\lambda_3-\lambda_1}
\,.
\end{equation}
All of this allows one to write $E_{\3Q}(\ell,h)$ in parametric form with the same parameters $v$ and $\lambda_3$ as before. 

Since we are interested in the situation when the string connected to $Q_3$ is long enough, we have to consider the behavior near $\lambda_3=1$. If so, then the expression for the energy takes the form 

\begin{equation}\label{EL-}
E_{\3Q}=\sigma \vert Q_3Y\vert
-
\g\sqrt{\s}
\Bigl(
{\cal I}(v)
-
2{\cal E}^-(\lambda_1,v)
-
3\k\frac{\ep^{-2v}}{\sqrt{v}}\,
\Bigr)
+
3c+o(1)\,,
\end{equation}
whereas the expressions for $\lambda_1$ and $\beta$ take the form 

\begin{equation}\label{lamdi2}
\lambda_1=-\text{ProductLog}\biggl(-v\,\ep^{-v}
\biggl[1-\frac{1}{4}\biggl(3\k(1+4v)\ep^{-3v}+
\Bigl[1-v^2\ep^{2(1-v)}\Bigr]^{\oh}\biggr)^2\,\biggr]^{-\oh}
\,\biggr)
\,,\qquad
\sin\beta=\frac{1}{2}\lambda_1\ep^{1-\lambda_1}
\,.
\end{equation}
Here the parameter $v$ runs from $\vit$ to $\vet$, where $\vet$ is a solution to the equation $\lambda_1(v)=1$. We denote this solution by $\vet$ because it is the same as that in the case of the equilateral triangle geometry. One way to see this is to rewrite the equation $\lambda_1(v)=1$ as  

\begin{equation}\label{vast}
	\k(1+4v)\ep^{-3v}+\Bigl[1-v^2\,\ep^{2(1-v)}\Bigr]^{\oh}=0
	\,.
\end{equation}

In the limit $v\rightarrow \vet$, the angle $\beta$ tends to $\frac{\pi}{6}$, and hence $Y$ tends to the Fermat point of the triangle.\footnote{$\beta(v)$ is an increasing function of $v$ which takes values on the interval $[0,\pi/6]$. Thus, for a general $v$ the point $Y$ is different from the Fermat point.} Moreover, as $\vert Q_1Y\vert /\vert Q_3Y\vert\rightarrow 1$ the isosceles triangle approaches the equilateral one. This is the IR limit of the $QQQ$ system such that all the quarks are away from each other. The energy is then given by 

\begin{equation}\label{Edi-IR}
E_{\3Q}=\sigma \vert Q_3Y\vert+\frac{2}{\sqrt{3}}\sigma\ell-3\g\sqrt{\s}\,I_{\3Q}+3c+o(1)
\,
\end{equation}
that is nothing but the Y-law \eqref{E3Q-etg-large} expressed in terms of $\vert Q_3Y\vert$ and $\ell$. 

Thus, we started from the diquark limit at $v=0$ and finally reached the equilateral triangle geometry at $v=\vet$ by separating the neighbor quarks $Q_1$ and $Q_2$.

\subsubsection{Disconnected configuration and string breaking}

For the decay mode under consideration the corresponding disconnected configuration is that of Figure \ref{etg} but with $Q_3$ placed at $y=h$. Under the assumption of non-interacting hadrons, its energy is the same sum of $E_{\QQq}$ and $E_{\Qqb}$. 

Now, we want to find a critical value of $\vert Q_3Y\vert$ by solving the equation $E_{\3Q}=E_{\QQq}+E_{\Qqb}$. That gives the string breaking distance 

\begin{equation}
\boldsymbol{\ell}_{\3Q}^{(1)}=\vert Q_3Y\vert_c
\,
\end{equation} 
meaning that for fixed $\ell$, the disconnected configuration dominates in the expectation value of the Wilson loop for $\vert Q_3Y\vert>\vert Q_3Y\vert_c$. Since both functions $E_{\3Q}$ and $E_{\QQq}$ are defined piecewise so that each subinterval has a size depending on a value of $\k$, for definiteness we restrict to $\k=-0.102$.\footnote{Results for other values of $\k$ are similar.} We also choose the parameter set $L$. A straightforward but somewhat tedious calculation, using the linear approximations \eqref{EL+} and \eqref{EL-}, shows that the equation $E_{\3Q}=E_{\QQq}+E_{\Qqb}$ is solved by 

\begin{equation}\label{l3Q}
	\boldsymbol{\ell}_{\3Q}^{(1)}=
	\begin{cases}
\boldsymbol{\ell}_{0}+\frac{2}{\ep\sqrt{\s}}
\bigl({\cal E}^+(\alpha',v')-{\cal E}^+(\alpha_1,v)\bigr)
\quad &\hspace{-0.35cm}\text {if}\quad 0\leq \ell\leq 0.248\,\text{fm}\,,
\\
\boldsymbol{\ell}_1+\frac{2}{\ep\sqrt{\s}}\bigl({\cal E}^+(\alpha',v')-{\cal E}^+(\alpha_1,v)\bigr)
\quad &\hspace{-0.35cm}\text {if}\quad 0.248\leq \ell\leq 0.779\,\text{fm}\,,
\\
	\boldsymbol{\ell}_1+\frac{2}{\ep\sqrt{\s}}\bigl({\cal E}^+(\alpha',v')-{\cal E}^-(\lambda,v)\bigr)
\quad &\hspace{-0.35cm} \text {if}\quad 0.779\leq \ell\leq 1.110\,\text{fm}\,,
\\	
\boldsymbol{\ell}_1+\frac{2}{\ep\sqrt{\s}}\bigl({\cal E}^-(\lambda',v')-{\cal E}^-(\lambda,v)\bigr)
	\quad &\hspace{-0.35cm}\text {if}\quad 1.110 \leq \ell\leq 1.577\,\text{fm}
	\,,
	\end{cases}
\end{equation}
with 
\begin{equation}\label{l0}
\begin{split}
\boldsymbol{\ell}_0=
\frac{1}{\ep\sqrt{\s}}
\Bigl(
2{\cal Q}(q)-{\cal Q}(v')
+
2\n\frac{\ep^{\oh q}}{\sqrt{q}}
+
3\k\Bigl(\frac{\ep^{-2v'}}{\sqrt{v'}}
-
\frac{\ep^{-2v}}{\sqrt{v}}\Bigr)
+
{\cal I}(v)
\Bigr)
\,,
\\
\boldsymbol{\ell}_1=
\frac{1}{\ep\sqrt{\s}}
\Bigl(
{\cal Q}(q)
+
\n\Bigl(\frac{\ep^{\oh q}}{\sqrt{q}}+\frac{\ep^{\oh v'}}{\sqrt{v'}}\Bigr)
+
3\k\Bigl(\frac{\ep^{-2v'}}{\sqrt{v'}}
-
\frac{\ep^{-2v}}{\sqrt{v}}\Bigr)
+
{\cal I}(v)
\Bigr)
\,.
\end{split}
\end{equation}
Here $v$ is the parameter for the $QQQ$ configuration. The function $\alpha_1(v)$ is defined by \eqref{anglesdi1} and $\lambda(v)$ by \eqref{lamdi2}. The parameter $v'$ refers to the $QQq$ configuration. The corresponding functions 
$\alpha'(v')$ and $\lambda'(v')$ are defined by \eqref{alpha1}, \eqref{alpha2},  and \eqref{lambda-QQq}. The base length is expressed by the two functions $\ell=\ell(v)$ and $\ell=\ell(v')$ which are defined in the previous subsection and Appendix D. Because of this, the parameters are not independent. Note that the value $\ell=1.577\,\text{fm}$ corresponds to $\boldsymbol{\ell}_{\3Q}^{(1)}=0.910\,\text{fm}$ obtained for the equilateral triangle geometry in Sec.III. 

\begin{figure}[H]
\centering
\includegraphics[width=7.25cm]{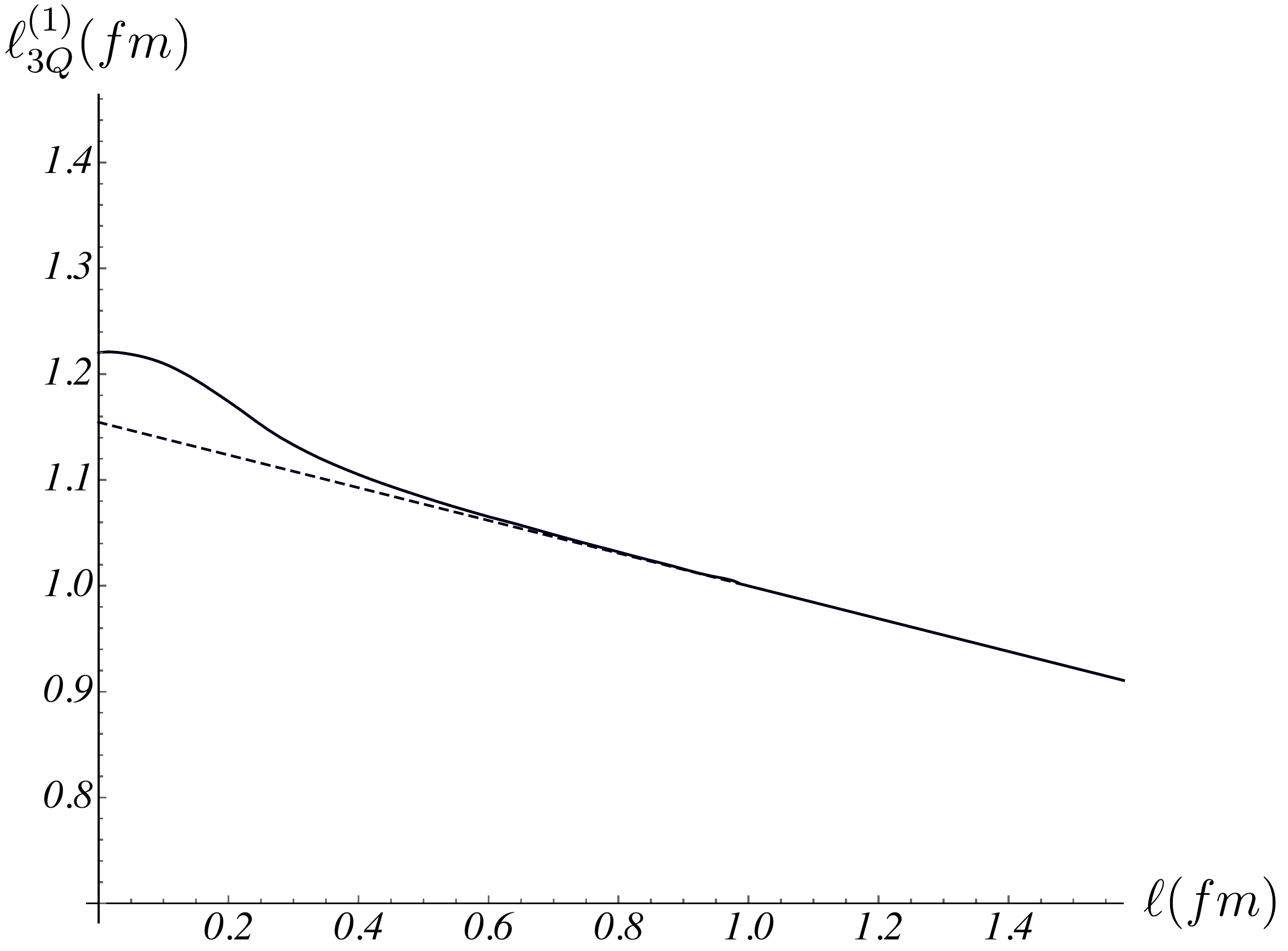}
\hspace{2cm}
\includegraphics[width=7.25cm]{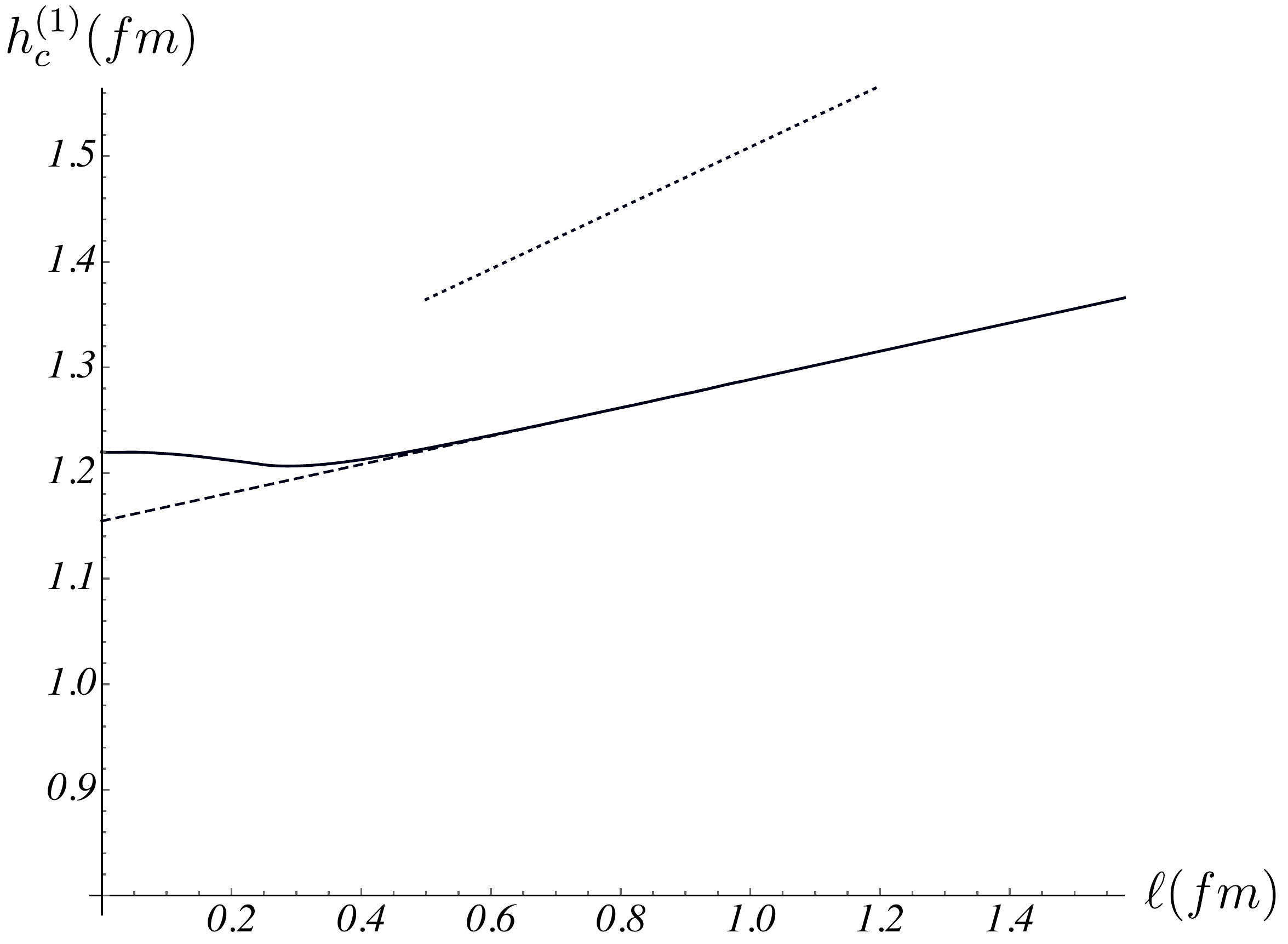}
\caption{{\small Left: $\boldsymbol{\ell}_{\3Q}^{(1)}$ as a function of $\ell$. Right: $h_c^{(1)}$ as a function of $\ell$. The dashed lines are the asymptotic behaviors obtained from  \eqref{Q3-large} and \eqref{hc-large}, and dotted line from \eqref{hc-universal}.}} 
\label{lhc}
\end{figure}

Given the parametric equations $\boldsymbol{\ell}_{\3Q}^{(1)}=\boldsymbol{\ell}_{\3Q}^{(1)}(v)$ and $\ell=\ell(v)$, we can eliminate the parameter and find $\boldsymbol{\ell}_{\3Q}^{(1)}=\boldsymbol{\ell}_{\3Q}^{(1)}(\ell)$ numerically. The result is presented in Figure \ref{lhc} on the left. We see that $\boldsymbol{\ell}_{\3Q}^{(1)}$ does decrease with increasing the base 
\noindent length of the triangle. 
 
There is certainly more that we could say about this geometry. The point is that for small enough $\ell$ the decay mode we are considering is dominant. This means that the ground state is determined in terms of two states: $QQQ$ and $QQq+Q\bar q$, and the scale $\boldsymbol{\ell}_{\3Q}^{(1)}$ is the only scale in question. One way to roughly estimate the upper bound for $\ell$ is to accept that the $QQq$ system becomes unstable for $\ell\gtrsim \boldsymbol{\ell}_{\QQb}$ \cite{a-QQq}. If so, then we expect that only one string breaks down if the separation between $Q_1$ and $Q_2$ (length of the triangle's base) is less than $1.22\,\text{fm}$. For larger separations, the other decay modes become relevant that makes the whole picture of string breaking quite complicated.      

In order to make comparison with other approaches easier, we introduce a critical height $h_c^{(1)}=\vert Q_3Y\vert_c+\oh\tan\beta\,\ell$. Similarly to $\boldsymbol{\ell}_{\3Q}^{(1)}$, we can also eliminate $v$ and find $h_c^{(1)}=h_c^{(1)}(\ell)$ numerically. The result of this is shown in Figure \ref{lhc} on the right. In contrast to $\boldsymbol{\ell}_{\3Q}^{(1)}$, the critical height behaves in a more complicated way. It is nearly $\boldsymbol{\ell}_{\QQb}$ for $\ell$ below $0.2\,\text{fm}$ and then, after a small decrease, starts to linearly increase with $\ell$ for $\ell$ above $0.3\,\text{fm}$. 

Now consider the diquark limit so that $\ell\rightarrow 0$. It is useful to write the first expression in \eqref{l3Q} as  

\begin{equation}\label{Q3s}
\boldsymbol{\ell}_{\3Q}^{(1)} =
\boldsymbol{\ell}_{\QQb}
+\frac{1}{\ep\sqrt{\s}}
\Bigl[
2{\cal E}^+(\alpha'_1,v')-2{\cal E}^+(\alpha_1,v)
+
3\k\Bigl(\frac{\ep^{-2v'}}{\sqrt{v'}}-\frac{\ep^{-2v}}{\sqrt{v}}\Bigr)
-
{\cal Q}(v')
-
\frac{1}{\sqrt{v}}
+
\int_0^{\sqrt{v}}\frac{du}{u^2}\bigl(\ep^{u^2}[1-u^4\ep^{2(1-u^2)}]^\oh-1\bigr)\Bigr]
\,,
\end{equation}
with $\boldsymbol{\ell}_{\QQb}$ given by \eqref{Lc-mes}. The terms inside the square brackets vanish as $\ell\rightarrow 0$. This immediately follows from the fact that this limit corresponds to $v\rightarrow 0$ and $v'\rightarrow 0$. If so, then in leading order $v=v'$ and $\cos\alpha_1=\cos\alpha'_1=\frac{\sqrt{3}}{2}(1-2\k-3\k^2)^{\frac{1}{2}}$.  So, one obtains $\boldsymbol{\ell}_{\3Q}^{(1)}=\boldsymbol{\ell}_{\QQb}$, as expected from quark-diquark symmetry.

The behavior for large enough $\ell$ is 

\begin{equation}\label{Q3-large}
\boldsymbol{\ell}_{\3Q}^{(1)}=\bigl(1-\tfrac{2}{\sqrt{3}}\bigr)\ell
+
\frac{1}{\ep\sqrt{\s}}
\Bigl(
{\cal Q}(q) +\n\frac{\ep^{\oh q}}{\sqrt{q}}+3I_{\3Q}-2I_{\QQq}
\Bigr)
\,,
\end{equation}
as it follows from \eqref{Edi-IR} and \eqref{EQQq-large}. For $\gamma=\frac{\pi}{3}$, where $\ell=\sqrt{3}\boldsymbol{\ell}_{\3Q}^{(1)}$, it reduces to the expression \eqref{Lc-etg} obtained for the equilateral triangle geometry. With the help of this expression, one finds the asymptotic behavior of the critical height 

\begin{equation}\label{hc-large}
h_c^{(1)}=\bigl(1-\tfrac{\sqrt{3}}{2}\bigr)\ell
+
\frac{1}{\ep\sqrt{\s}}
\Bigl(
{\cal Q}(q) +\n\frac{\ep^{\oh q}}{\sqrt{q}}+3I_{\3Q}-2I_{\QQq}
\Bigr)
\,.
\end{equation}

We conclude this discussion with one more remark. If one assumes that the string breaking distance is universal, i.e. $\boldsymbol{\ell}_{3Q}^{(1)}=\boldsymbol{\ell}_{\QQb}$, then the behavior of the critical heigh for large $\ell$ is   

\begin{equation}\label{hc-universal}
h_c^{(1)}=\tfrac{1}{2\sqrt{3}}\ell+\boldsymbol{\ell}_{\QQb}
\,. 	
\end{equation}
This suggests that $h_c^{(1)}$ increases more rapidly for large values of $\ell$ (see Figure \ref{lhc}). Hopefully, it will be possible eventually to determine $h_c^{(1)}(\ell)$ reliably by computer simulations. 

\subsection{Collinear geometry}

Now let us consider the second type of geometry. It can be obtained from the diquark limit in which $Q_1$ sits on top of $Q_2$ by moving $Q_1$ along the axis connecting $Q_2$ and $Q_3$ in the direction away from $Q_3$. As a result, we get the collinear geometry. 
\subsubsection{Connected string configurations}

\begin{figure}[htbp]
\centering
\includegraphics[width=7cm]{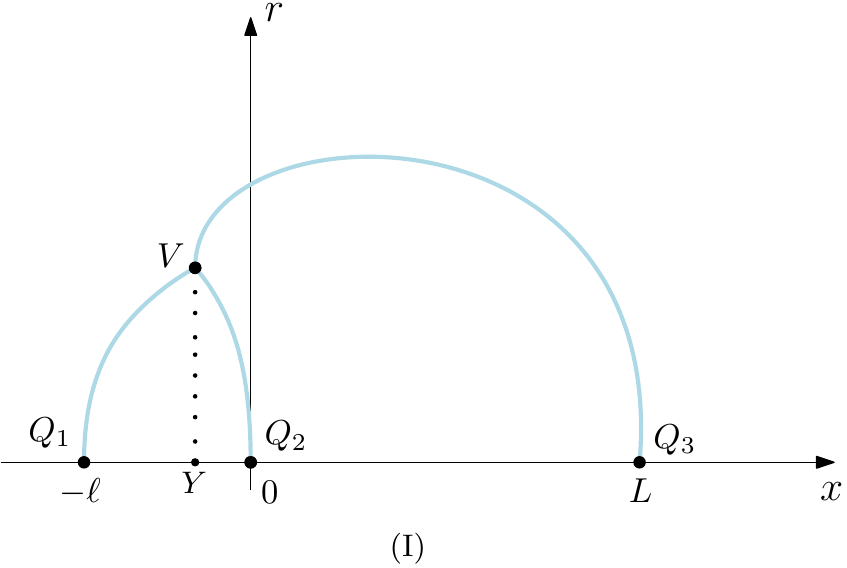}
\hspace{2cm}
\includegraphics[width=7cm]{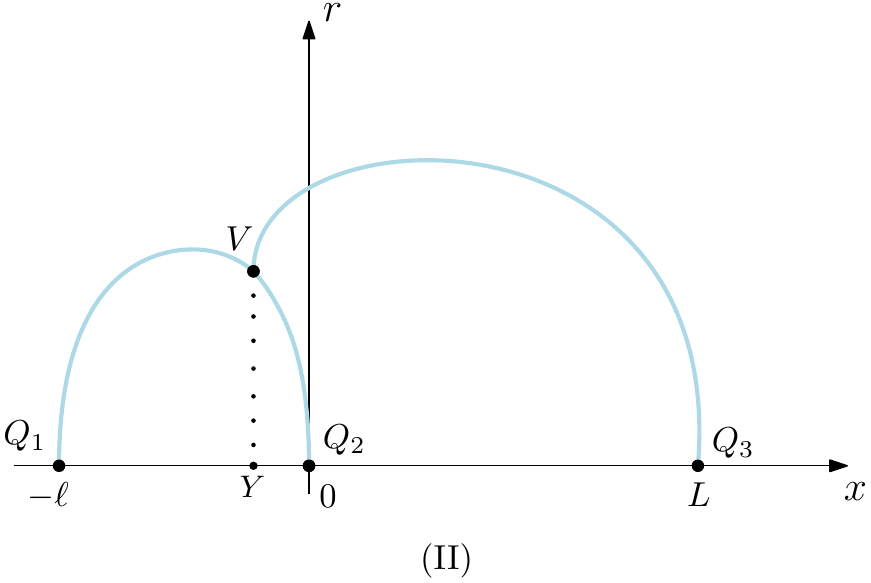}
\caption{{\small Typical collinear configurations if $L\geq\ell$. Left: The case $\alpha_1\geq0$. Right: The case $\alpha_1\leq 0$.}}
\label{colc}
\end{figure}

For this case, the connected configurations were studied in \cite{a-3q}. These are shown in Figure \ref{colc}.
An important observation which is inferred from moving $Q_1$ is that at some point the string connected to it changes the shape from that of Figure \ref{strings} on the left to that on the right. In the meantime the others keep their shape intact. This change corresponds to a flip of the sign of $\alpha_1$. 

We begin with configuration I describing the diquark limit. Using the formulas of Appendix B for the case $\alpha\geq 0$, we immediately deduce that

\begin{equation}\label{l-cg}
	\ell=\frac{1}{\sqrt{\s}}\bigl({\cal L}^+(\alpha_1,v)+{\cal L}^+(\alpha_2,v)\bigr)
	\,,\quad
	L=\frac{1}{\sqrt{\s}}\bigl({\cal L}^-(\lambda_3,v)-{\cal L}^+(\alpha_2,v)\bigr)
	\,,
\end{equation}
and the energy of the configuration is  

\begin{equation}\label{Ecg}
E_{\3Q}=\g\sqrt{\s}\Bigl(
\sum_{i=1}^2 {\cal E}^+(\alpha_i,v)\,+{\cal E}^-(\lambda_3,v)
	+
	3\k\frac{\ep^{-2v}}{\sqrt{v}}
\Bigr)+3c
\,.
\end{equation}

In addition, the angles $\alpha_1$ and $\alpha_2$ can be defined implicitly as functions of $v$ and $\lambda_3$ by the force balance equations \eqref{fbv-cg}. We have 

\begin{equation}\label{alphas2}
\cos\alpha_1=\cos\alpha_2+\frac{v}{\lambda_3}\ep^{\lambda_3-v}
\,,\quad
\Bigl[1-\Bigl(\cos\alpha_2+\frac{v}{\lambda_3}\ep^{\lambda_3-v}\Bigr)^2\Bigr]^{\oh}
+
\Bigl[1-\cos^2{}\hspace{-1mm}\alpha_2\Bigr]^{\oh}
-
\Bigl[1-\frac{v^2}{\lambda_3^2}\ep^{2(\lambda_3-v)}]\Bigr]^{\oh}
=
3\k(1+4v)\ep^{-3v}
\,.
\end{equation}
Thus, the energy of the configuration is given in parametric form by the equations $E_{\3Q}=E_{\3Q}(v,\lambda_3)$, $\ell=\ell(v,\lambda_3)$, and $L=L(v,\lambda_3)$. The parameters take values on the interval $[0,1]$ and obey the inequality $v\leq \lambda_3$.

The behavior of this configuration for small $\ell$ was studied in \cite{a-3q}. But for our purposes, we don't need all the results of this study. Only the leading term in the expansion of $\ell(v)$ near $v=0$ matters. It is given by Eq.\eqref{l-v}, exactly as in the case of the triangle geometry.  
 
On the other hand, we do need to know what happens if the string ended at $Q_3$ is long enough. Taking the limit $\lambda_3\rightarrow 1$ in \eqref{Ecg} and \eqref{alphas2}, we get 

\begin{equation}\label{Ecg-large}
E_{\3Q}=\sigma\vert Q_3Y\vert
	+
	\g\sqrt{\s}
	\Bigl(\sum_{i=1}^2 {\cal E}^+(\alpha_i,v)
	+
	3\k\frac{\ep^{-2v}}{\sqrt{v}}
-
{\cal I}(v)
\Bigr)+3c
\,
\end{equation}
and 
\begin{equation}\label{alphas3}
\cos\alpha_1=\cos\alpha_2+v\ep^{1-v}
\,,\quad
\Bigl[1-\Bigl(\cos\alpha_2+v\ep^{1-v}\Bigr)^2\Bigr]^{\oh}
+
\Bigl[1-\cos^2{}\hspace{-1mm}\alpha_2\Bigr]^{\oh}
-
\Bigl[1-v^2\ep^{2(1-v)}]\Bigr]^{\oh}
=
3\k(1+4v)\ep^{-3v}
\,.
\end{equation}
The angles $\alpha_1$ and $\alpha_2$ depend only on $v$ which varies from $0$ to $\tilde v$, where $\tilde v$ is a solution to 

\begin{equation}\label{vc}
\Bigl[1-\bigl(1-v\ep^{1-v}\bigr)^2\Bigr]^{\oh}
-
\Bigl[1-v^2\ep^{2(1-v)}]\Bigr]^{\oh}
=
3\k(1+4v)\ep^{-3v}
\,.
\end{equation}
This parameter value corresponds to the transition between the configurations of Figure \ref{colc} because $\alpha_1(\tilde v)=0$. 

It is straightforward to obtain the corresponding expressions for configuration II. The only modifications of the above formulas are due to the string which changes its shape. Using the formulas of Appendix B for the case $\alpha\leq 0$, we find 

\begin{equation}\label{l-cg2}
	\ell=\frac{1}{\sqrt{\s}}\bigl({\cal L}^-(\lambda_1,v)+{\cal L}^+(\alpha_2,v)\bigr)
	\,,\quad
	L=\frac{1}{\sqrt{\s}}\bigl({\cal L}^-(\lambda_3,v)-{\cal L}^+(\alpha_2,v)\bigr)
	\,,
\end{equation}
and 
\begin{equation}\label{Ecg2}
E_{\3Q}=
	\g\sqrt{\s}\Bigl(
	\sum_{i=1,3}{\cal E}^-(\lambda_i,v)+{\cal E}^+(\alpha_2,v)
	+
3\k\frac{\ep^{-2v}}{\sqrt{v}}
\Bigr)+3c
\,.
\end{equation}
With the help of \eqref{v-lambda}, the force balance equations \eqref{fbv-cg} can be written as 

\begin{equation}\label{alphal2}
\begin{split}
\cos\alpha_2=&v\ep^{-v}\Bigl(\frac{\ep^{\lambda_1}}{\lambda_1}-\frac{\ep^{\lambda_3}}{\lambda_3}\Bigr)
\,, \\
\Bigl[1-v^2\ep^{-2v}\Bigl(\frac{\ep^{\lambda_1}}{\lambda_1}-\frac{\ep^{\lambda_3}}{\lambda_3}\Bigr)^2
\Bigr]^{\oh}
-\Bigl[1-\frac{v^2}{\lambda_1^2}&\ep^{2(\lambda_1-v)}\Bigr]^{\oh}
-
\Bigl[1-\frac{v^2}{\lambda_3^2}\ep^{2(\lambda_3-v)}]\Bigr]^{\oh}
=
3\k(1+4v)\ep^{-3v}
\,.
\end{split}
\end{equation}
In general, we are unable to explicitly express one parameter as a function of two others in the last equation. In practice it is convenient to choose $v$ and $\lambda_3$ as independent parameters and then solve this equation for $\lambda_1$ numerically. 

In the limit $\lambda_3\rightarrow 1$, the expression for the energy takes the form

\begin{equation}\label{Ecg2-1}
E_{\3Q}=\sigma\vert Q_3Y\vert+
	\g\sqrt{\s}\Bigl(
	{\cal E}^-(\lambda_1,v)+{\cal E}^+(\alpha_2,v)
	+
	3\k\frac{\ep^{-2v}}{\sqrt{v}}
	-{\cal I}(v)
\Bigr)
+
3c
\,,
\end{equation}
and the equations \eqref{alphal2} take the form 

\begin{equation}\label{alphal2-1}
	\begin{split}
	\cos\alpha_2=&v\ep^{1-v}\Bigl(\frac{\ep^{\lambda_1-1}}{\lambda_1}-1\Bigr)
	\,,\\
\Bigl[1-v^2\ep^{2(1-v)}\Bigl(\frac{\ep^{\lambda_1-1}}{\lambda_1}-1 \Bigr)^2
\Bigr]^{\oh}
-\Bigl[1-\frac{v^2}{\lambda_1^2}&\ep^{2(\lambda_1-v)}\Bigr]^{\oh}
-
\Bigl[1-v^2\ep^{2(1-v)}]\Bigr]^{\oh}
=
3\k(1+4v)\ep^{-3v}
\,.
\end{split}
\end{equation}
The last equation now determines $\lambda_1$ as a function of $v$. The parameter $v$ runs from $\tilde v$ to $\bar v$. The upper bound satisfies $\lambda_1(\bar v)=1$. It is the same as that for the symmetric collinear configuration studied in Sec.II. Thus, starting from the diquark limit, we reach the symmetric configuration by separating the heavy quarks $Q_1$ and $Q_2$.

\subsubsection{Disconnected configuration and string breaking}

As before, the relevant disconnected configuration is of type shown in Figure \ref{col-d} on the left. There is a little difference between those configurations. The quark $Q_3$ is now at $x=L$, with $L\geq\ell$. However, this has no effect on the total energy of the configuration under the assumption of non-interacting hadrons. So both configurations have the same energy $E_{\QQq}+E_{\Qqb}$.

Having understood the string configurations, we can now find the critical value of $\vert Q_3Y\vert$ that defines the string breaking distance $\boldsymbol{\ell}_{\3Q}^{(1)}=\vert Q_3Y\vert_c$ by solving the equation $E_{\3Q}=E_{\QQq}+E_{\Qqb}$. So, at fixed $\ell$ the disconnected configuration dominates in the expectation value of the Wilson loop for $\vert Q_3Y\vert>\vert Q_3Y\vert_c$. As in the previous subsection, we take the parameter set $L$ and $\k=-0.102$. Using the linear approximations \eqref{Ecg-large} and \eqref{Ecg2-1} makes the things easy. So, we get 

\begin{equation}\label{l3Q-cl}
	\boldsymbol{\ell}_{\3Q}^{(1)}=
	\begin{cases}
\boldsymbol{\ell}_{0}
+
\frac{1}{\ep\sqrt{\s}}
\bigl({2\cal E}^+(\alpha',v')-{\cal E}^+(\alpha_1,v)-
{\cal E}^+(\alpha_2,v)\bigr)
\quad &\hspace{-0.35cm}\text {if}\quad 0\leq \ell\leq 0.087\,\text{fm}\,,
\\
\boldsymbol{\ell}_{0}
+
\frac{1}{\ep\sqrt{\s}}
\bigl({2\cal E}^+(\alpha',v')-{\cal E}^-(\lambda_1,v)-
{\cal E}^+(\alpha_2,v)\bigr)
\quad &\hspace{-0.35cm}\text {if}\quad 
0.087\leq \ell\leq 0.248\,\text{fm}
\,,
\\
\boldsymbol{\ell}_{1}
+
\frac{1}{\ep\sqrt{\s}}
\bigl({2\cal E}^+(\alpha',v')-{\cal E}^-(\lambda_1,v)-
{\cal E}^+(\alpha_2,v)\bigr)
\quad &\hspace{-0.35cm}\text {if}\quad 
0.248\leq \ell\leq 1.110\,\text{fm}
\,,
\\	
\boldsymbol{\ell}_{1}
+
\frac{1}{\ep\sqrt{\s}}
\bigl({2\cal E}^-(\lambda',v')-{\cal E}^-(\lambda_1,v)-
{\cal E}^+(\alpha_2,v)\bigr)
\quad &\hspace{-0.35cm}\text {if}\quad 
1.110 \leq \ell \lesssim 1.110+\epsilon\,\text{fm}
	\,,
	\end{cases}
\end{equation}
where $\boldsymbol{\ell}_{0}$ and $\boldsymbol{\ell}_{1}$ are the same two functions as in \eqref{l3Q}. The parameters $v$ and $v'$ refer respectively to the connected configurations for $QQQ$ and $QQq$. These are not independent because of the relation $\ell=\ell(v)=\ell'(v')$. The functions $\alpha_1(v)$, $\alpha_2(v)$, and $\lambda_1(v)$ are defined implicitly by \eqref{alphas3} and \eqref{alphal2-1}. As before in \eqref{l3Q}, $\alpha'(v')$ and $\lambda'(v')$ are defined by \eqref{alpha1}, \eqref{alpha2},  and \eqref{lambda-QQq}. The last interval in \eqref{l3Q-cl} is really small, with $\epsilon\approx0.0003$. The string breaking distance tends to that of the symmetric configuration as $\ell$ approaches the upper bound.

\begin{figure}[htbp]
\centering
\includegraphics[width=7.25cm]{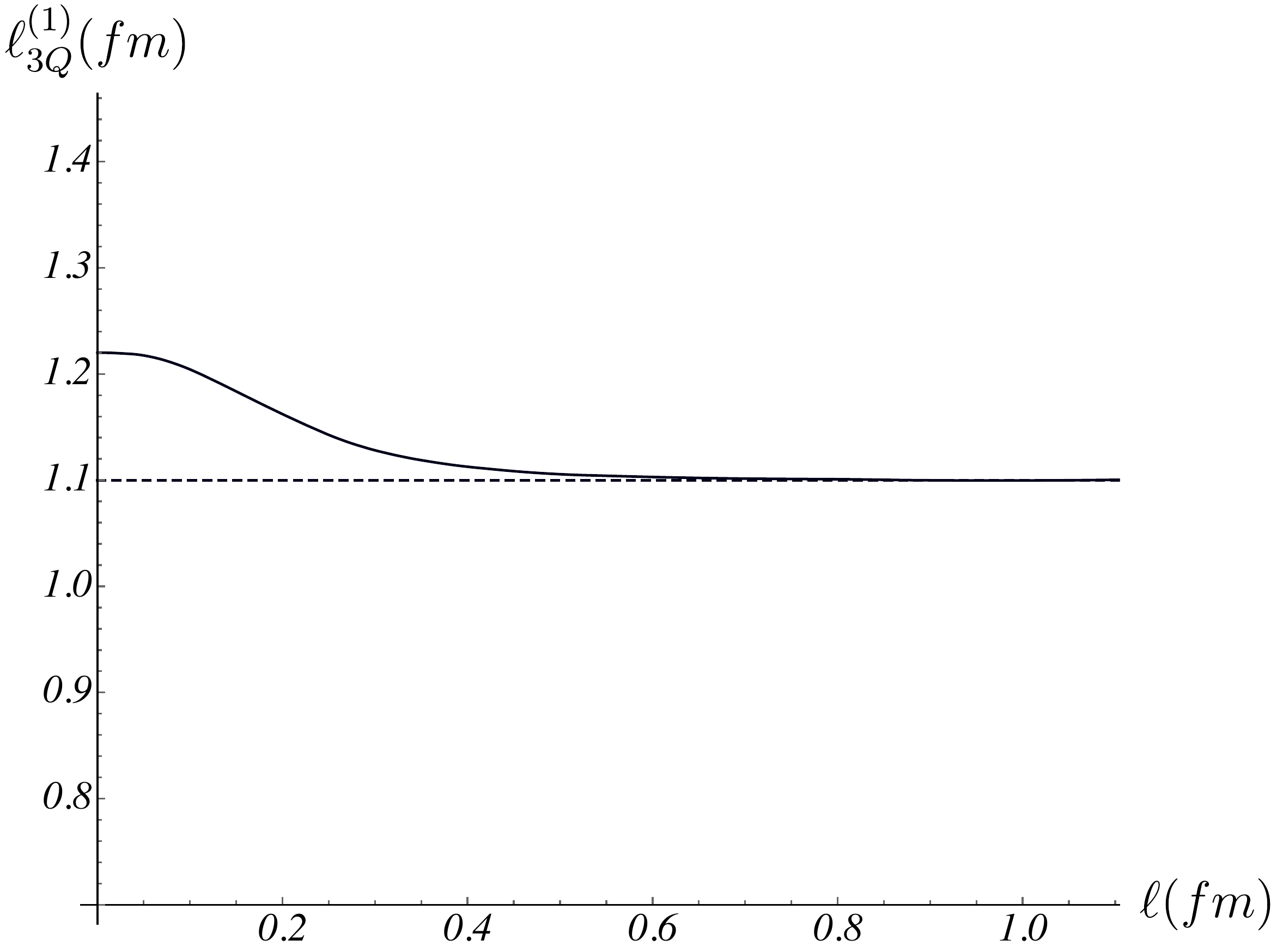}
\hspace{2cm}
\includegraphics[width=7.25cm]{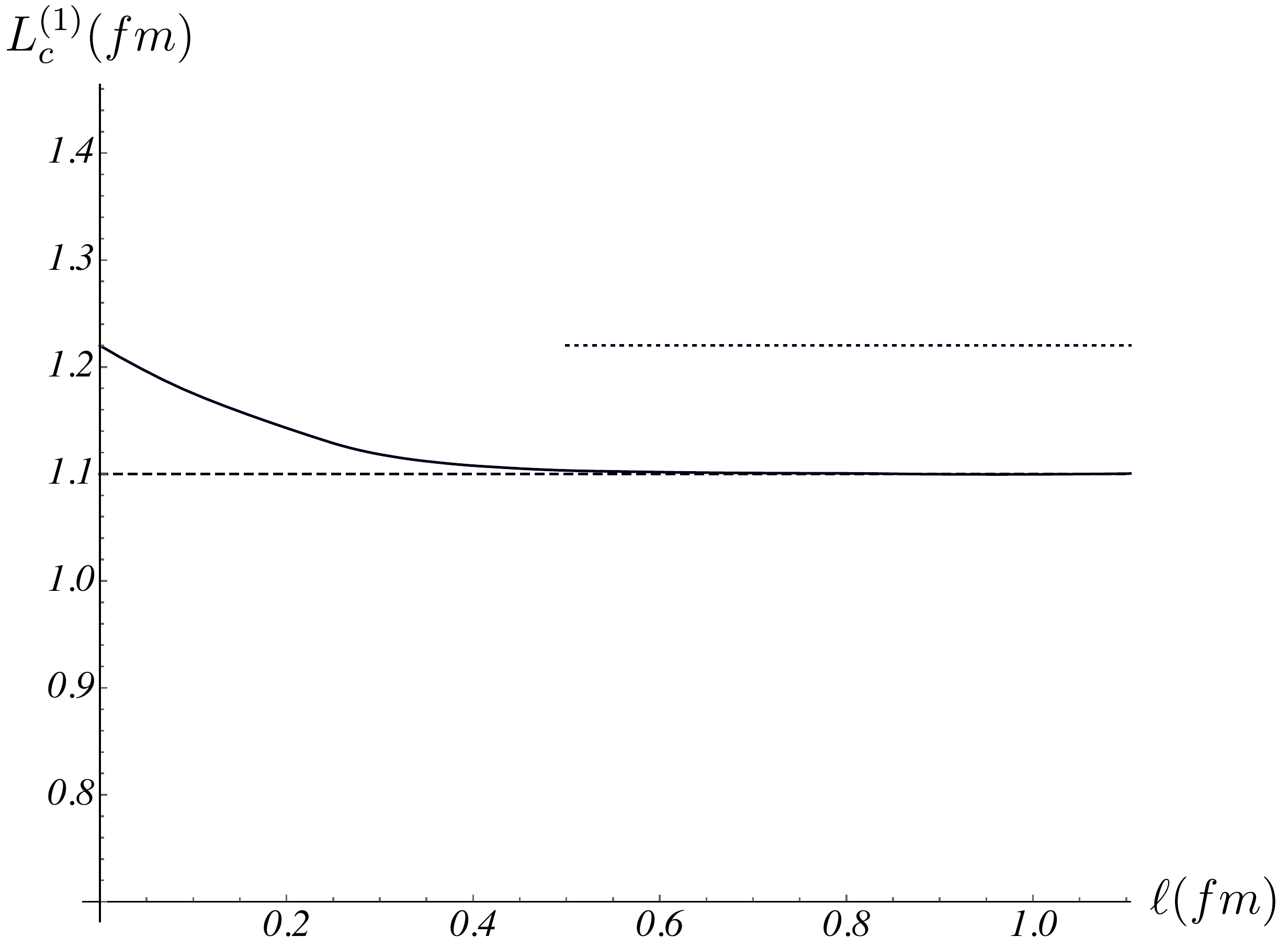}
\caption{{\small $\boldsymbol{\ell}_{\3Q}^{(1)}$ and $L_c^{(1)}$ as functions of $\ell$. The dashed horizontal lines represent the value $1.110\,\text{fm}$ and dotted line the value $1.22\,\text{fm}$. The former corresponds to the value we found in Sec.III and the latter to the value of $\boldsymbol{\ell}_{\QQb}$ from \cite{bulava}.}}
\label{cgLc}
\end{figure}

One can numerically eliminate the remaining parameter from the parametric equations $\boldsymbol{\ell}_{\3Q}^{(1)}=\boldsymbol{\ell}_{\3Q}^{(1)}(v)$ and $\ell=\ell(v)$, and find a corresponding function $\boldsymbol{\ell}_{\3Q}^{(1)}(\ell)$. The result is shown in Figure \ref{cgLc} on the left. We see that the string 
breaking distance coincides with $\boldsymbol{\ell}_{\QQb}$ at $\ell=0$ and then decreases with increasing the separation between the neighbor quarks. A notable feature is that for separations larger than $0.5\,\text{fm}$ it becomes almost the same as that for the symmetric collinear geometry. 

Before proceeding further, we pause here to say that for small $\ell$ the decay mode $QQQ\rightarrow QQq+Q\bar q$ is dominant. So, as before for the isosceles triangle geometry, the ground state is determined in terms of two states and $\boldsymbol{\ell}_{\3Q}^{(1)}$ is the only scale. The present case shows that the estimate of the upper bound for $\ell$ via the instability of $QQq$ is not so accurate. It gives $1.22\,\text{fm}$, but $\ell$ is limited to $1.100\,\text{fm}$ which corresponds to the symmetric case. Thus, another assumption is needed to determine the upper bound more accurately.

Now we introduce a critical length $L_c^{(1)}=\vert Q_3Y\vert_c-\vert Q_2Y\vert$ that can be useful in looking for possible ways to make comparisons with other non-perturbative approaches. The meaning of $L_c^{(1)}$ is that at a given value of $\ell$, the disconnected configuration dominates for lengths longer than $L_c^{(1)}$. Like above, we can eliminate $v$ and find $L_c^{(1)}=L_c^{(1)}(\ell)$. The result is presented in Figure \ref{cgLc}. We see that so defined $L_c^{(1)}$ behaves quite similarly to $\boldsymbol{\ell}_{\3Q}^{(1)}$. 

The small $\ell$ behavior of $\boldsymbol{\ell}_{\3Q}^{(1)}$ is determined through the same steps as before. First, we bring the first expression in \eqref{l3Q-cl} into the following form 

\begin{equation}\label{Lc-cgs}
\boldsymbol{\ell}_{\3Q}^{(1)}=
\boldsymbol{\ell}_{\QQb}+\frac{1}{\ep\sqrt{\s}}
\Bigl[2{\cal E}^+(\alpha',v')-\sum_{i=1}^2{\cal E}^+(\alpha_i,v)
+
3\k\Bigl(\frac{\ep^{-2v'}}{\sqrt{v'}}-\frac{\ep^{-2v}}{\sqrt{v}}\Bigr)
-{\cal Q}(v')-
\frac{1}{\sqrt{v}}
+
\int_0^{\sqrt{v}}\frac{du}{u^2}\bigl(\ep^{u^2}[1-u^4\ep^{2(1-u^2)}]^\oh-1\bigr)\Bigr]
\,.
\end{equation}
Then, from the equation $\ell(v)=\ell'(v')$ we obtain that in leading order $v=v'$ and $\alpha_1=\alpha_2=\alpha'$. This implies that the terms inside the square brackets cancel each other out as $v$ and $v'$ go to zero. Thus, we get the desired result $\boldsymbol{\ell}_{\3Q}^{(1)}=\boldsymbol{\ell}_{\QQb}$.

To see what happens for large $\ell$, we consider the double limit $\lambda_1\rightarrow 1$ and $\lambda'\rightarrow 1$ such that $\frac{\ln (1-\lambda_1)}{\ln (1-\lambda')}\rightarrow 1$. The requirement implies that $\frac{\ell}{L}\rightarrow 1$.\footnote{This follows from the asymptotic expansion \eqref{lE-large}.} Thus this limit gives rise to the symmetric case. A simple calculation shows that the last expression in \eqref{l3Q-cl} does reduce to that of \eqref{lc1-col}. The latter is shown in the Figure by the dashed horizontal lines.     

Finally, let us mention that if one assumes that the string breaking distance is universal, then the critical length is simply $L_c^{(1)}=\boldsymbol{\ell}_{\QQb}$. In other words, so defined critical length is a constant, which is independent of $\ell$. It is shown in the Figure by the dotted line. 

\section{More detail on the special cases}
\renewcommand{\theequation}{5.\arabic{equation}}
\setcounter{equation}{0}

\begin{figure}[htbp]
\centering
\includegraphics[width=6.25cm]{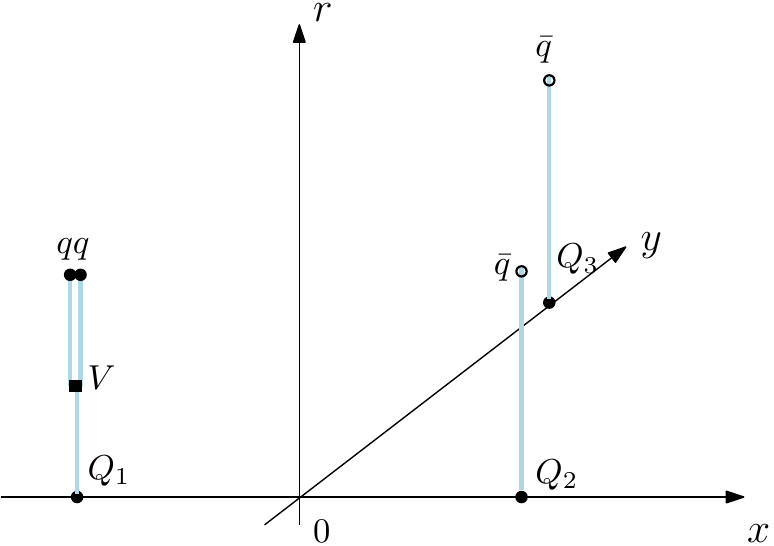}
\hspace{1.5cm}
\includegraphics[width=6.25cm]{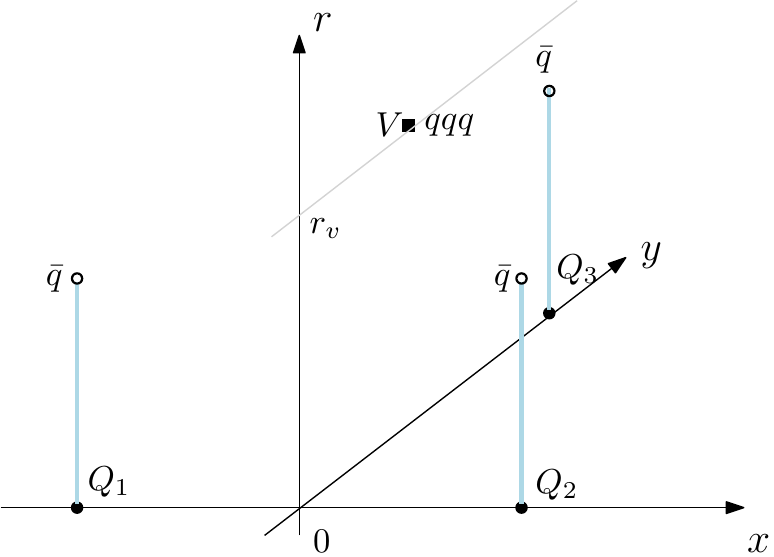}
\caption{{\small Static string configurations for non-interacting hadrons. The ordering does not matter. Left: A heavy-light baryon $Qqq$ and two heavy-light mesons $Q\bar q$. Right: A light baryon $qqq$ and three heavy-light mesons.}}
\label{dconf23}
\end{figure}

For completeness, we include the analysis of the remaining decay modes in \eqref{decay} which respectively contain two and three pairs of light quarks each. To actually compute the energies of non-interacting hadrons within the gauge/string duality formalism, one would have to consider disconnected string configurations  such as those sketched in Figure \ref{dconf23}. There is an important subtlety that arises when one tries to
construct a static configuration for a heavy-light baryon $Qqq$. The analysis of Eq.\eqref{v||} shows that this equation has a solution in the interval $[0,1]$ if and only if $-\frac{\ep^3}{15}\leq\k\leq\km$ \cite{a-QQq}. Clearly, the phenomenologically motivated value $\k=-0.102$ is out of the interval. This is one of the limitations of using the model of Sec.II. We take the upper bound as it is most close to $\k=-0.102$ and leads to the exact solution $\vp=\frac{1}{12}$ which makes the calculations simpler. 

First, let us consider the configuration sketched in the left panel. It describes the products of the baryon decay into $Qqq+2Q\bar q$. Under the assumption of non-interacting hadrons, the energy of the configuration is the sum of energies of individual hadrons. These energies can be read off from Eqs.\eqref{EQqb} and \eqref{EQqq}. So, we have   

\begin{equation}\label{EQqq+2mes}
	E_{\Qqq}+2E_{\Qqb}=\g\sqrt{\s}\Bigl(4{\cal Q}(q)-{\cal Q}(\vp)
	+4\n\frac{\ep^{\oh q}}{\sqrt{q}}
	+3\k\frac{\ep^{-2\vp}}{\sqrt{\vp}}\,\Bigr)+3c
	\,,
\end{equation}
where $q$ and $\vp$ are the solutions of \eqref{q} and \eqref{v||}, respectively. 

The right panel represents a string configuration for the products of the baryon decay into $qqq+3Q\bar q$. The novelty here is a light baryon that has not been discussed in the literature. In the static limit, it looks like the light quarks sit on top of the baryon vertex. The total action is therefore given by 

\begin{equation}\label{Sqqq}
S=\frac{T}{r_v}\Bigl(\tau_v\ep^{-2\s\rv^2}+3\m \ep^{\frac{\s}{2}\rv^2}\Bigr)
\,.
\end{equation}
By varying it with respect to $\rv$, we get

\begin{equation}\label{v3}
\n (1-v)+\k(1+4v)\ep^{-\frac{5}{2}v}=0
\,. 
\end{equation}
This equation says that the gravitational force acting on the vertex is equilibrated by that acting on the quarks. As a result, the baryon is at rest. The rest energy is then  

\begin{equation}\label{Eqqq}
	E_{\qqq}=3\g\sqrt{\frac{\s}{v_{\qqq}}}\bigl(\k\ep^{-2v_{\qqq}}+\n\ep^{\oh v_{\qqq}}\Bigr)
	\,,
\end{equation}
where $v_{\qqq}$ is a solution of equation \eqref{v3} on the interval $[0,1]$. Combining this with the expression \eqref{EQqb}, we find the energy of the configuration 

\begin{equation}\label{Eqqq+3mes}
	E_{\qqq}+3E_{\Qqb}=3\g\sqrt{\s}\Bigl(
	{\cal Q}(q)+\n\frac{\ep^{\oh q}}{\sqrt{q}}
	+\k\frac{\ep^{-2v_{\qqq}}}{\sqrt{v_{\qqq}}}+\n\frac{\ep^{\oh v_{\qqq}}}{\sqrt{v_{\qqq}}}
	\,\Bigr)+3c
	\,.
\end{equation}

\subsection{Equilateral triangle geometry} 

\begin{figure}[htbp]
\centering
\includegraphics[width=7.5cm]{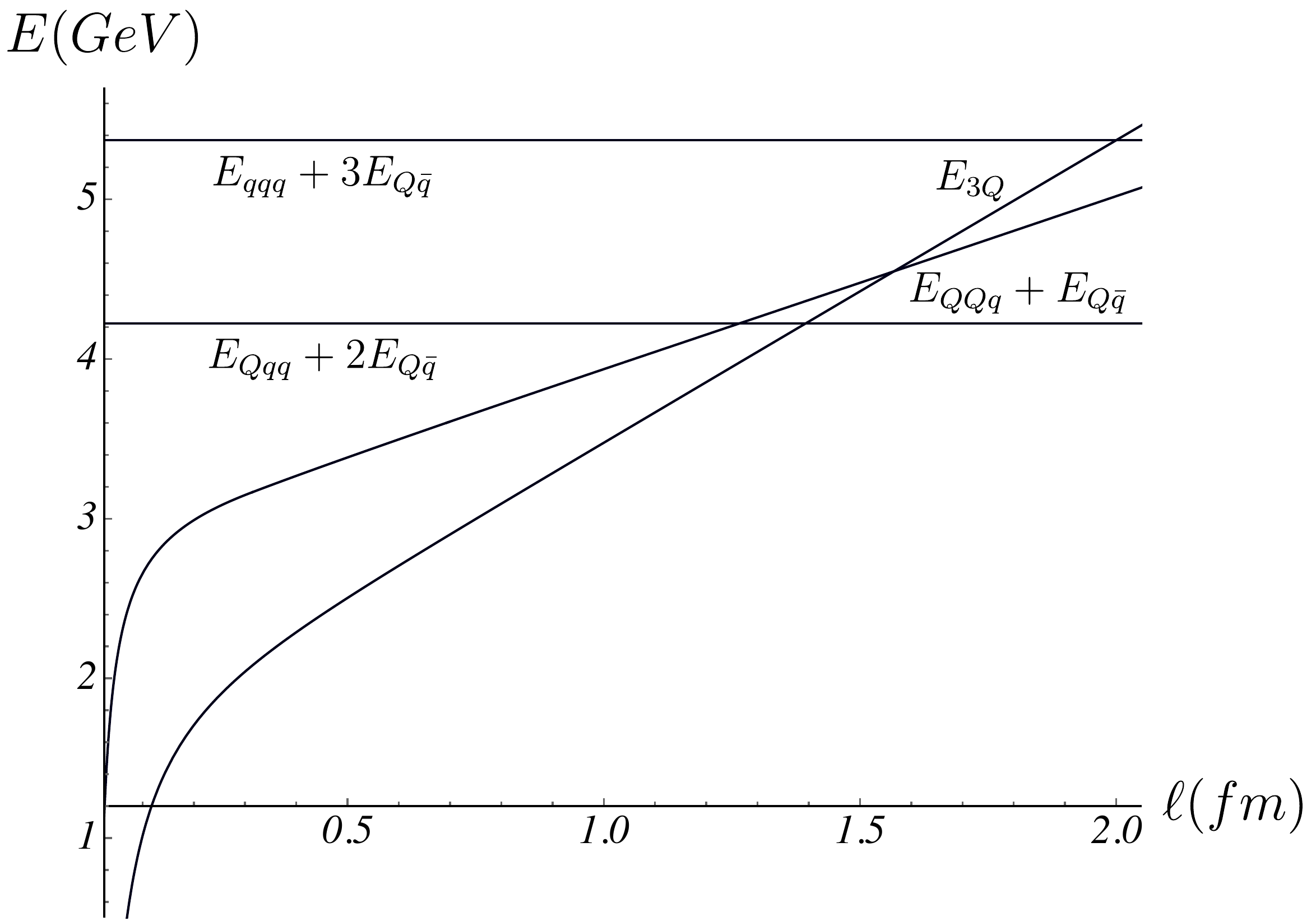}
\hspace{1.5cm}
\includegraphics[width=7.5cm]{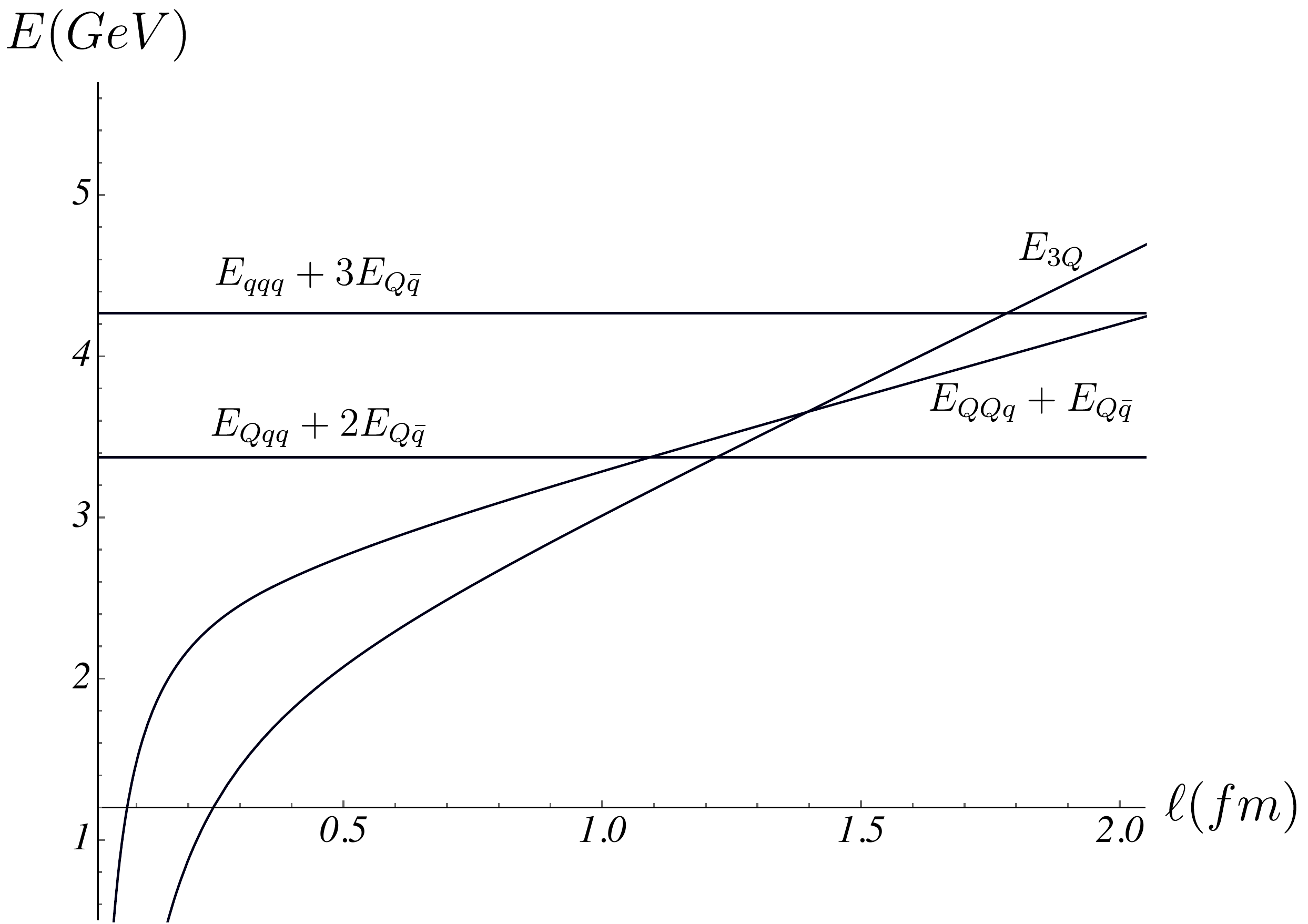}
\caption{{\small Various $E$ vs $\ell$ plots for the equilateral triangle geometry, shown in the left panel for $L$ and in the right panel for $P$. Here $\k=\km$ and $c=0.623\,\text{GeV}$.}}
\label{V0}
\end{figure}

With all the diagonal elements of the model Hamiltonian at our disposal, we can see a more complete picture of string breaking. We consider first the equilateral triangle geometry. To this end, in Figure \ref{V0} we plot the diagonal elements as a function of $\ell$.  
The most interesting observation is that the three-quark potential (ground state energy) is, in fact, determined in terms of only two diagonal elements: $E_{\3Q}$ and $E_{\Qqq}+2E_{\Qqb}$. Thus, the string breaking distance should be defined from the equation $E_{\3Q}=E_{\Qqq}+2E_{\Qqb}$, i.e. the $QQQ\rightarrow Qqq+2Q\bar q$ mode is dominant. If so, then with the help of \eqref{E3Q-etg-large} and \eqref{EQqq+2mes}, we find that 

\begin{equation}\label{lc2}
\boldsymbol{\ell}_{\3Q}^{(2)}=\frac{1}{3\ep\sqrt{\s}}
\Bigl(
4{\cal Q}(q) - {\cal Q}(\vp)+4\n\frac{\ep^{\oh q}}{\sqrt{q}}+3\k\frac{\ep^{-2\vp}}{\sqrt{\vp}}
+
3I_{\3Q}
\Bigr)
\,.
\end{equation}
Here the superscript (2) indicates that the string breaking distance refers to the decay mode $QQQ\rightarrow Qqq+2Q\bar q$. 

Quite apart from the above observation, it is to be noticed that the decay $Qqq\rightarrow qqq+Q\bar q$ is energetically forbidden because of $E_{qqq}+E_{\Qqb}>E_{Qqq}$. In the string model we are using, this means that the string stretched along the radial direction (see the right panel of Figure \ref{Qqb}) does not break down. 

Let us make a simple estimate of $\boldsymbol{\ell}_{\3Q}^{(2)}$. At $\k=\km$, for $L$ and $P$ we get correspondingly

\begin{equation}\label{l3Q(2)}
\boldsymbol{\ell}_{\3Q}^{(2)}=0.804\,\text{fm}
\,,\qquad
\boldsymbol{\ell}_{\3Q}^{(2)}=0.700\,\text{fm}
\,.	
\end{equation}
These values are smaller than those obtained for $\boldsymbol{\ell}_{\3Q}^{(1)}$ in Sec.III, as also seen from the plots. In addition, we get\footnote{Curiously, both values are close to $\frac{2}{3}$.} 

\begin{equation}\label{lc2/lcQQb}
\frac{\boldsymbol{\ell}_{\3Q}^{(2)}}{\boldsymbol{\ell}_{\QQb}}=0.659
\,,\,\,
\qquad
\frac{\boldsymbol{\ell}_{\3Q}^{(2)}}{\boldsymbol{\ell}_{\QQb}}=0.652
\,.
\end{equation}
Thus, for the equilateral triangle geometry, the string theory prediction is actually that the three quark potential is determined in terms of the energies $E_{\3Q}$ and $E_{\Qqq}+2E_{\Qqb}$ with a single scale which characterizes the string breaking effect.
\subsection{Symmetric collinear geometry}

\begin{figure}[htbp]
\centering
\includegraphics[width=7.5cm]{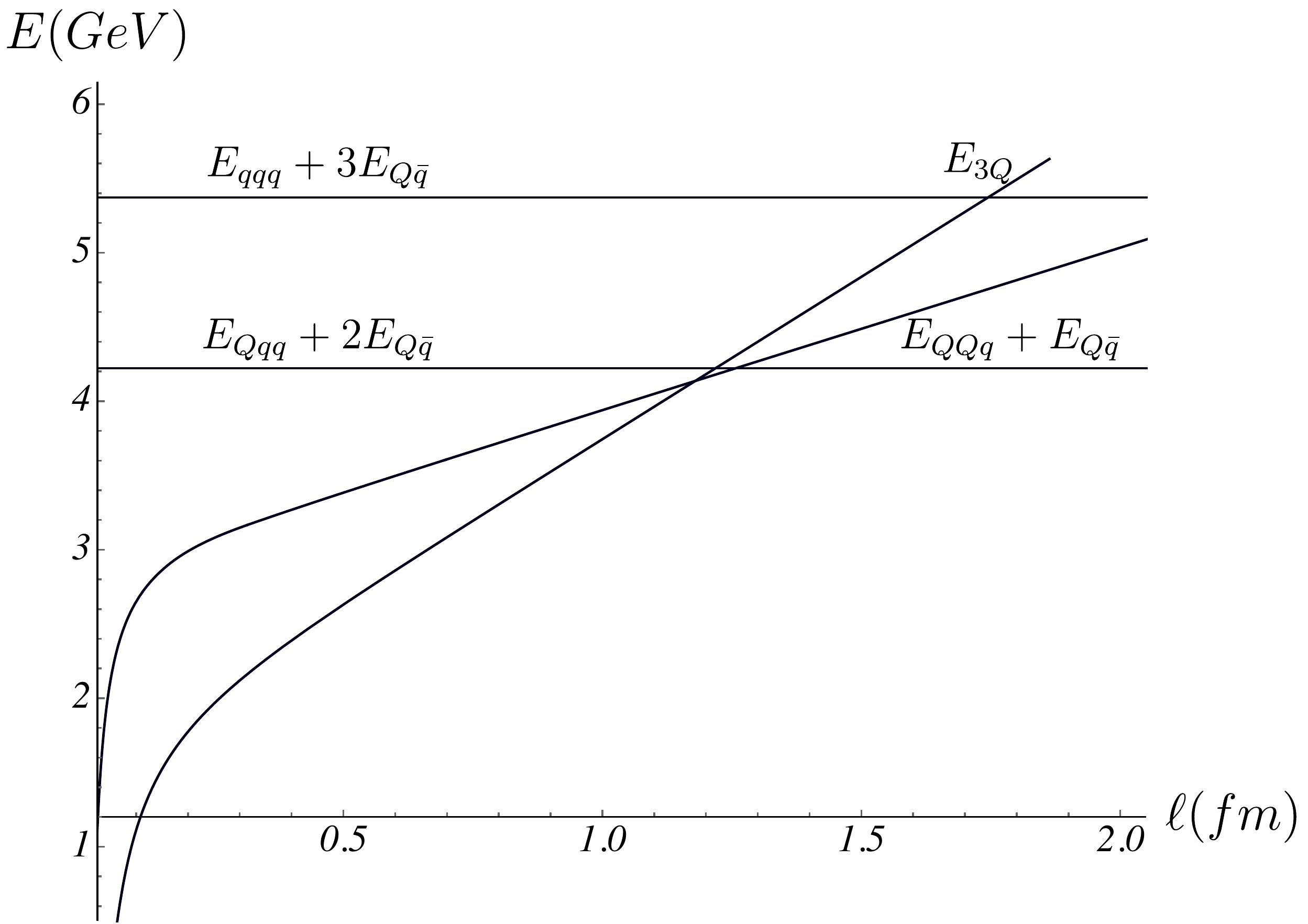}
\hspace{1.5cm}
\includegraphics[width=7.5cm]{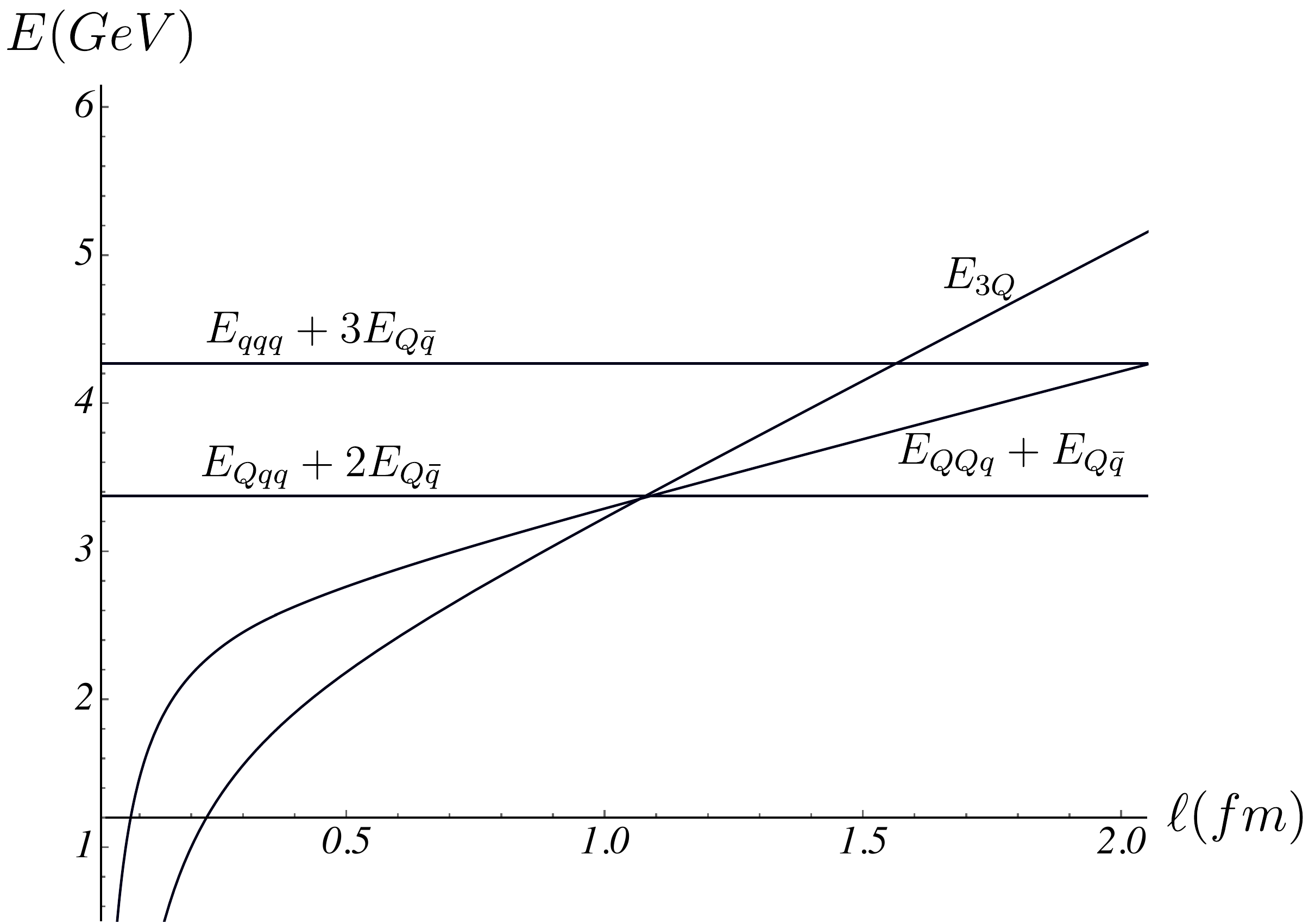}
\caption{{\small Various $E$ vs $\ell$ plots for the symmetric collinear geometry. The notation here is the same as in Figure \ref{V0}.}}
\label{V0C}
\end{figure}

Similarly, we can analyze the symmetric collinear geometry. Under the assumption of non-interacting hadrons, the energies of the two remaining disconnected configurations are given by Eqs.\eqref{EQqq+2mes} and \eqref{Eqqq+3mes}. In Figure \ref{V0C} we plot the diagonal elements of ${\cal H}$ as a function of $\ell$. The noticeable difference from the previous case is that 
the ground state energy is determined in terms of three diagonal elements: $E_{\3Q}$, $E_{\QQq}+E_{\Qqb}$, and $E_{\Qqq}+2E_{\Qqb}$.\footnote{Because the difference between $\boldsymbol{\ell}_{\3Q}^{(1)}$ and $\boldsymbol{\ell}_{QQq}$ is of order $0.003\,\text{fm}$ for $P$, it is not visible in the right panel.}  This fact suggests that a scale set by $E_{\QQq}(\ell)+E_{\Qqb}=E_{\Qqq}+2E_{\Qqb}$ is relevant. In fact, this scale was recently discussed in the context of the $QQq$ system \cite{a-QQq}. It is given by  

\begin{equation}\label{lc-Ex}
\boldsymbol{\ell}_{\QQq} =\frac{3}{\ep\sqrt{\s}}
\biggl(
{\cal Q}(q)-\frac{1}{3}{\cal Q}(\vp) +\n\frac{\ep^{\oh q}}{\sqrt{q}}+\k\frac{\ep^{-2\vp}}{\sqrt{\vp}}+\frac{2}{3}I_{\QQq}
\biggr)
\,.
\end{equation}
The estimates give $\boldsymbol{\ell}_{\QQq}=1.257\,\text{fm}$ for the parameter set $L$ and $\boldsymbol{\ell}_{\QQq}=1.073\,\text{fm}$ for P. These values are very close to those for $\boldsymbol{\ell}_{\QQb}$. All of this suggests that the $QQQ$ system undergoes a cascade of decays. At small $\ell$ it looks like a triply heavy baryon. Then, at $\ell=\boldsymbol{\ell}_{\3Q}^{(1)}$ this baryon decays into a double-heavy baryon and a heavy light meson. Finally, at $\ell=\boldsymbol{\ell}_{\QQq}$ the double heavy baryon decays into a pair of heavy-light baryon and meson. Thus, for the symmetric collinear geometry, the string model predicts that the potential is determined in terms of the energies $E_{\3Q}$, $E_{\QQq}+E_{\Qqb}$, and $E_{\Qqq}+2E_{\Qqb}$ with two scales which characterize the string breaking effects.

Finally, let us note that small enough deformations of these special geometries will give similar pictures of string breaking. 

\section{Concluding remarks}
\renewcommand{\theequation}{6.\arabic{equation}}
\setcounter{equation}{0}
(i) In this paper, we have initiated a study of the phenomenon of string breaking in the three quark system. Our findings could be an early indication that this phenomenon is much more complex than one could naively expect based on the universality of the string breaking distance. The main funding is that the string breaking distance is not universal. Moreover, there are special geometries, such as the symmetric collinear geometry and its small deformations, which allow more than one characteristic scale. 

(ii) Going back to the question of universality, we see from Eq. \eqref{E13} that in the IR limit the energy $E_{\3Q}$ is\footnote{This is valid if the largest angle of the triangle $Q_1Q_2Q_3$ is at most $2\pi/3$.}

\begin{equation}
	E_{\3Q}=\sigma\sum_{i=1}^3\vert Q_i Y\vert\, -
	3\g\sqrt{\s}\,{\cal I}(\vet)+3c+o(1)
	\,,
\end{equation}
with $\vet$ a solution of Eq.\eqref{vast}. In this limit, $Y$ coincides with the Fermat point of the triangle \cite{a-3q0}. The coefficient $\sigma$ is the string tension which is universal in the sense that it is the same for all the known examples of heavy quark systems. The two remaining constant terms combine to $C_{\text{IR}}=3c-3\g\sqrt{\s}{\cal I}(\vet)$. Using the fact that in UV limit (at small separations between the quark sources) the constant term in the expansion of $E_{\3Q}$ is simply $C_{\text{UV}}=3c$, we get

\begin{equation}\label{CIR}
	C_{\text{UV}}-C_{\text{IR}}=3\g\sqrt{\s}{\cal I}(\vet)
	\,.
\end{equation}
This difference being positive is also scheme and geometry independent. The latter means that it is universal and not limited to the particular case of a triangle if its largest angle is at most $\frac{2}{3}\pi$. This statement is an extension of that in \cite{a-3q}, which was formulated for the equilateral triangle.  

(iii) Although from the ten-dimensional perspective the five-dimensional string model used as an example throughout this paper is oversimplified, it allows us to get the results on the $QQQ$ system analytically and uses only a few model parameters. There are, however, some limitations and caveats for this model, as for any effective model. Let us mention a couple of them. First, the light quarks are incorporated along the lines of the hadro-quarkonium picture so that $\rq$ denotes an averaged position of a light quark or a center of a corresponding quark cloud. The model certainly is far from accounting for all the features of $u$ and $d$ quarks. Second, it is not clear how to compute the off-diagonal elements of the model Hamiltonians and thus reach the level of understanding the quark-antiquark system in lattice gauge theory. 

\begin{acknowledgments}
 We would like to thank M. Catillo, P. de Forcrand, M. K. Marinkovi\'c, and P. Weisz for discussions and reading the manuscript. We also thank the Arnold Sommerfeld Center for Theoretical Physics for hospitality. This research is supported by Russian Science Foundation grant 20-12-00200 in association with Steklov Mathematical Institute.
\end{acknowledgments}

\appendix
\section{Notation and definitions}
\renewcommand{\theequation}{A.\arabic{equation}}
\setcounter{equation}{0}

In all Figures throughout the paper, heavy and light quarks (antiquarks) are denoted by $Q$ and $q\,(\bar q)$, and a baryon vertex and its projection onto the boundary of five-dimensional space by $V$ and $Y$. A square indicates that a light quark sits on top of a vertex. We assume that all strings are in the ground state. So, these strings are represented by curves without cusps, loops, etc. When not otherwise noted, we usually set light quarks (antiquarks) at $r=\rq$ and a vertex at $r=\rv$. For convenience, we introduce two dimensionless variables $q=\s\rq^2$ and $v=\s\rv^2$. They take values on the interval $[0,1]$ and show how far from the soft-wall these objects are. 

To make formulas more compact, we introduce a set of basic functions. The motivation for that comes from the analysis of a static string in Appendix B. 

The non-negative function ${\cal L}^+$ is defined by the integral

\begin{equation}\label{fL+}
{\cal L}^+(\alpha,x)=\cos\alpha\sqrt{x}\int^1_0 du\, u^2\, \ep^{x (1-u^2)}
\Bigl[1-\cos^2{}\hspace{-1mm}\alpha\, u^4\ep^{2x(1-u^2)}\Bigr]^{-\frac{1}{2}}
\,,
\qquad
0\leq\alpha\leq\frac{\pi}{2}\,,
\qquad 
0\leq x\leq 1
\,.
\end{equation}
This function vanishes if $\alpha=\frac{\pi}{2}$ or $x=0$. It diverges at $(0,1)$. 

The other non-negative function ${\cal L}^-$ is defined as follows

\begin{equation}\label{fL-}
{\cal L}^-(y,x)=\sqrt{y}
\biggl(\,
\int^1_0 du\, u^2\, \ep^{y(1-u^2)}
\Bigl[1-u^4\,\ep^{2y(1-u^2)}\Bigr]^{-\frac{1}{2}}
+
\int^1_
{\sqrt{\frac{x}{y}}} 
du\, u^2\, \ep^{y(1-u^2)}
\Bigl[1-u^4\,\ep^{2y(1-u^2)}\Bigr]^{-\frac{1}{2}}
\,\biggr)
\,,
\quad
0\leq x\leq y\leq 1
\,.
\end{equation}
It vanishes at the origin and diverges at $y=1$. At $y=x$, ${\cal L}^-$ reduces to ${\cal L}^+$ with $\alpha=0$.  

It is also useful to define two other functions. The first is given by 

\begin{equation}\label{fE+}
{\cal E}^+(\alpha,x)=\frac{1}{\sqrt{x}}
\int^1_0\,\frac{du}{u^2}\,\biggl(\ep^{x u^2}
\Bigl[
1-\cos^2{}\hspace{-1mm}\alpha\,u^4\ep^{2x (1-u^2)}
\Bigr]^{-\frac{1}{2}}-1-u^2\biggr)
\,,
\qquad
0\leq\alpha\leq\frac{\pi}{2}\,,
\qquad 
0\leq x\leq 1
\,
\end{equation}
and the second by 

\begin{equation}\label{fE-}
{\cal E}^-(y,x)=\frac{1}{\sqrt{y}}
\biggl(
\int^1_0\,\frac{du}{u^2}\,
\Bigl(\ep^{y u^2}\Bigl[1-u^4\,\ep^{2y(1-u^2)}\Bigr]^{-\frac{1}{2}}
-1-u^2\Bigr)
+
\int^1_{\sqrt{\frac{x}{y}}}\,\frac{du}{u^2}\,\ep^{y u^2}
\Bigl[1-u^4\,\ep^{2y(1-u^2)}\Bigr]^{-\frac{1}{2}}
\biggr) 
\,,
\,\,\,
0\leq x\leq y\leq 1
\,.
\end{equation}
The function ${\cal E}^+$ diverges at $x=0$ and $(0,1)$, whereas the function ${\cal E}^-$ at $(0,0)$ and $y=1$. Just like for the ${\cal L}$'s, ${\cal E}^-$ reduces to ${\cal E}^+$ at $y=x$.

A special case of ${\cal E}^+$ is obtained by taking $\alpha=\frac{\pi}{2}$. In this case, the integral can be performed explicitly, yielding 

\begin{equation}\label{Q}
{\cal E}^+(\tfrac{\pi}{2},x)={\cal Q}(x)=\sqrt{\pi}\text{erfi}(\sqrt{x})-\frac{\ep^x}{\sqrt{x}}
\,.
\end{equation}
Here $\text{erfi}(x)$ is the imaginary error function. 

Finally, we define a function

\begin{equation}\label{I}
	{\cal I}(x)=I_0
-
\int_{\sqrt{x}}^1\frac{du}{u^2}\ep^{u^2}\Bigl[1-u^4\ep^{2(1-u^2)}\Bigr]^{\frac{1}{2}}\,
\,,
\qquad
I_0=\int_0^1\frac{du}{u^2}\Bigl(1+u^2-\ep^{u^2}\Bigl[1-u^4\ep^{2(1-u^2)}\Bigr]^{\frac{1}{2}}\Bigr)
\,,\qquad
0\leq x\leq 1
\,,
\end{equation}
which appears in the limit of sufficiently elongated strings. For the constant $I_0$, a simple numerical calculation gives $I_0\approx 0.751$.

\section{A static string with fixed endpoints}
\renewcommand{\theequation}{B.\arabic{equation}}
\setcounter{equation}{0}

In this Appendix we give a summary of the results on a static Nambu-Goto string in the curved geometry \eqref{metric}. These results provide the grounds for building string configurations of Secs. III and IV. Most of the material can be found in \cite{a-3q} whose conventions we generally follow.\footnote{The only new result presented here is a calculation of the constant term in the expansion of the energy of long strings.} So, we can be relatively brief.

\begin{figure}[htbp]
\centering
\includegraphics[width=5.5cm]{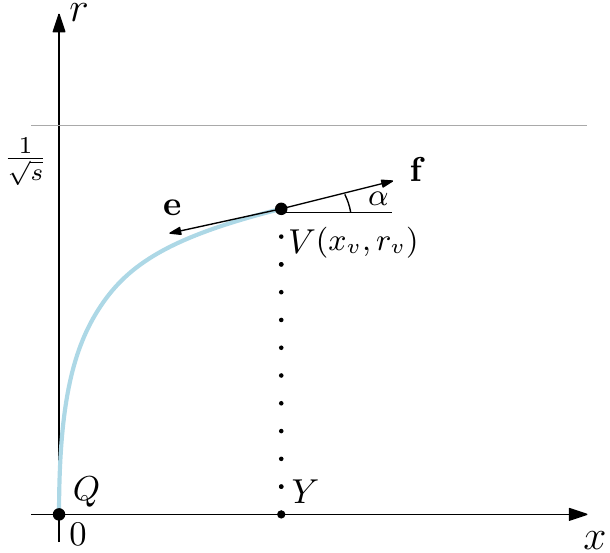}
\hspace{2.5cm}
\includegraphics[width=5.5cm]{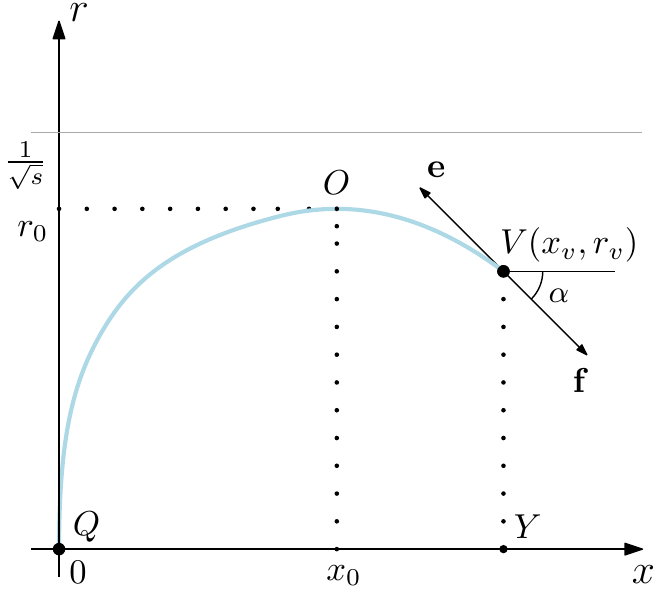}
\caption{{\small A static string stretched between two points, one of which lies on the boundary of space. The two forces exerted at point V are depicted by the arrows. $\alpha$ is a tangent angle. The string does not cross the soft-wall located at $r=1/\sqrt{\s}$. Left: The case $\alpha\geq 0$. Right: The case $\alpha\leq 0$. Here $O$ is a turning point.}}
\label{strings}
\end{figure}

To proceed, choose the static gauge $\xi^1=t$, giving for the Nambu-Got action 

\begin{equation}\label{NG-s}
S_{\text{\tiny NG}}=\frac{T}{2\pi\alpha'}\int_0^1 d\xi^2\,\sqrt{\gamma^{(2)}}
\,,
\end{equation}
with $T=\int_0^T dt$. Consider now a string stretched between two points $Q$ and $V$ in the $xr$-plane, as shown in Figure \ref{strings}. This implies the boundary conditions

\begin{equation}\label{string-bc}
x(0)=0\,,\quad x(1)=x_v\,,\quad r(0)=0\,,\quad r(1)=r_v
\,.
\end{equation}
In this case, the Nambu-Goto action takes the form

\begin{equation}\label{string-NG}
S_{\text{\tiny NG}}=\g T\int_{0}^{1} d\xi^2\,w(r)\sqrt{x'{}^2+r'{}^2}
\,,
\quad
\text{with}
\quad
w(r)=\frac{\ep^{\s r^2}}{r^2}
\,.
\end{equation}
A prime stands for a derivative with respect to $\xi^2$. 

Since in what follows we set infinitely heavy quarks at boundary points, we do not vary the action with respect to the position of point $Q$. The variation of the action is therefore 

\begin{equation}\label{variationNG}
\delta S_{\text{\tiny NG}}=\g T
\biggl[\int_{0}^{1} d\xi^2\,
\biggl(
\partial_r w\,\sqrt{x'{}^2+r'{}^2}-\Bigl(\frac{wr'}{\sqrt{x'{}^2+r'{}^2}}\Bigr)'\,
\biggr)\delta r
-
\Bigl(\frac{wx'}{\sqrt{x'{}^2+r'{}^2}}\Bigr)'\delta x 
+
w(\rv)\cos\alpha\, \delta x_v 
+
w(\rv)\sin\alpha\,\delta\rv 
\,
\biggr]
\,,
\end{equation}
where $\partial_r=\frac{\partial}{\partial r}$. From this formula one can find 
the string tension at point $V$.\footnote{For static configurations, the energy is simply $E=S/T$, while the components of $\mathbf{e}$ are $\mathbf{e}_x=-\delta E/\delta x_v$ and $\mathbf{e}_r=-\delta E/\delta\rv$.} It comes out to be 

\begin{equation}\label{e}
	\mathbf{e}=-\g w(\rv)\bigl(\cos\alpha,\,\sin\alpha\bigr)
	\,.
\end{equation}
There is one very simple but useful observation to add at this point. The magnitude of $\mathbf{e}$ equals $\|\mathbf{e}\|=\g w(\rv)$. This clarifies the role of the fifth dimension in describing strings with different tensions \cite{amp}.

In order to keep the string at rest, some external force must be acting on it. We exert such a force at point $V$, as shown in Figure \ref{strings}. The force balance equation is then 

\begin{equation}\label{fbe}
	\mathbf{e}+\mathbf{f}=0
	\,.
\end{equation}
Thus, given the string tension, the external force is $\mathbf{f}=-\mathbf{e}$.

For further analysis it is convenient to completely fix the gauge by choosing $\xi^2=x$. In that case, solving the Euler-Lagrange equation for $x$ gives 

\begin{equation}\label{Int}
I=\frac{w(r)}{\sqrt{1+(\partial_x r)^2}}
\,,
\end{equation}
with $I=const$. It is easy to see that $I$ is nothing else but the first integral for the equation for $r$ \cite{malda}. For future reference, we note that  

\begin{equation}\label{IntV}
I=w(\rv)\cos\alpha
\,
\end{equation}
at the endpoint $V$.
\subsection{The case $\alpha\geq 0$}

We begin with the case of positive $\alpha$ sketched in the left panel of Figure \ref{strings}. Combining \eqref{Int} and \eqref{IntV}, we get a differential equation $w(\rv)\cos\alpha=w(r)/\sqrt{1+(\partial_x r)^2}$ which can be further integrated over $x$ and $r$ to yield a string length along the $x$-axis 

\begin{equation}\label{l+}
\vert QY\vert=\frac{1}{\sqrt{\s}}{\cal L}^+(\alpha,v)
\,,
\end{equation}
where ${\cal L}^+$ and $v$ are defined earlier in Appendix A.

To compute the string energy, we first use \eqref{Int} to explicitly express the integrand in terms of $r$ and then find that  

\begin{equation}\label{e+}
E_{R}=\frac{S_R}{T}=\g\sqrt{\frac{\s}{v}}\int^1_{\sqrt{\tfrac{\s}{v}}\epsilon}\,\frac{du}{u^2}\,\ep^{v u^2}
\Bigl[1-\cos^2{}\hspace{-1mm}\alpha\,u^4\ep^{2v(1-u^2)}\Bigr]^{-\frac{1}{2}}
\,.
\end{equation}
Here the ultraviolet divergence at $r=0$ is regularized by placing a lower bound on $r$ so that $r\geq\epsilon$. This expression behaves for $\epsilon\rightarrow 0$ as 

\begin{equation}\label{e+R}
E_R=\frac{\g}{\epsilon}+E+O(\epsilon)\,.
\end{equation}
After subtracting the $\tfrac{1}{\epsilon}$ term and letting $\epsilon=0$, we arrive at

\begin{equation}\label{E+}
E=\g\sqrt{\s}\,{\cal E}^+(\alpha,v)+c
\,.
\end{equation}
where the function ${\cal E}^+$ is defined by \eqref{fE+} and $c$ is a normalization constant. 

If we set $\alpha=\frac{\pi}{2}$ in \eqref{E+}, we get a special case:

\begin{equation}\label{E|}
E=\g\sqrt{\s}{\cal Q}(v)+c
\,,
\end{equation}
with the function ${\cal Q}$ defined in \eqref{Q}. This formula gives the energy of a string stretched along the $r$-axis.  

Thus, the string energy is given in parametric form by $E=E(v)$ and $\vert QY\vert=\vert QY\vert(v)$. The parameter takes values on the interval $[0,1]$. The string becomes infinitely long at $v=1$, when its endpoint $V$ touches the soft-wall.

\subsection{The case $\alpha\leq0$}

It is straightforward to extend the above analysis to the case of negative $\alpha$, sketched in the right panel of Figure \ref{strings}. The only novelty, which applies to all expressions below, is that there are two contributions: one comes from the interval $[0,x_0]$ and the other from the interval $[x_0,x_v]$. This is so because one can think of a string with $\alpha<0$ as two strings with $\alpha=0$ glued at the turning point. 

Taking $I=w(r_0)$ and integrating the differential equation $w(r_0)=w(r)/\sqrt{1+(\partial_x r)^2}$ over $x$ and $r$, we get for a string length along the $x$-axis

\begin{equation}\label{l-}
\vert QY\vert=\frac{1}{\sqrt{\s}}{\cal L}^-(\lambda,v)
\,,
\end{equation}
where ${\cal L}^-$ is the function defined in Appendix A and $\lambda$ is a dimensionless parameter defined as $\lambda=\s r_0^2$. It shows how close to the soft-wall the turning point is. Clearly, at $\lambda=v$ the above expression reduces to that of \eqref{l+} with $\alpha=0$. This value also gives the lower bound $\frac{1}{\sqrt{\s}}{\cal L}^-(v,v)\leq \vert QY\vert$. 
 
As before, one can compute the string energy by first expressing the integrand in terms of $r$ and then imposing the short-distance cutoff on $r$. This gives

\begin{equation}\label{e-}
E_{R}=\g\sqrt{\frac{\s}{\lambda}}
\biggl(
\int^1_{\sqrt{\tfrac{\s}{\lambda}}\epsilon}\,\frac{du}{u^2}\,\ep^{\lambda u^2}
\Bigl[1-u^4\,\ep^{2\lambda (1-u^2)}\Bigr]^{-\frac{1}{2}}
+
\int^1_{\sqrt{\frac{v}{\lambda}}}\,\frac{du}{u^2}\,\ep^{\lambda u^2}
\Bigl[1-u^4\,\ep^{2\lambda (1-u^2)}\Bigr]^{-\frac{1}{2}}\biggr)
\,.
\end{equation}
Near $\epsilon=0$ it behaves like that in \eqref{e+R}. So, subtracting the $\frac{1}{\epsilon}$ and then letting $\epsilon=0$ yields 

\begin{equation}\label{E-}
E= \g\sqrt{\s}\,{\cal E}^-(\lambda,v)+c
\,.
\end{equation}
Here ${\cal E}^-$ is defined by \eqref{fE-} and $c$ is the same normalization constant as in \eqref{E+}. The last guarantees that the string energy is a continuous function of $\alpha$.

For future reference, it will be useful to explicitly express $\lambda$ in terms of $v$ and $\alpha$. This can be done by using the first integral \eqref{Int}. Equating its values at $O$ and $V$, one finds

\begin{equation}\label{v-lambda}
\frac{\ep^{\lambda}}{\lambda}=\frac{\ep^{v}}{v}\cos\alpha
\,.
\end{equation}
From this, $\lambda$ can be expressed in terms of $v$ and $\alpha$ as 

\begin{equation}\label{lambda}
\lambda=-\text{ProductLog}(-v\,\ep^{-v}/\cos\alpha)
\,.
\end{equation}
The ProductLog function is the principal solution for $w$ in $z=w\,\ep^w$ \cite{wolf}.

To make estimates of the string breaking distance, we will need to know the behavior of $E$ for large $\vert QY\vert$, or in other words what happens when the string approaches the soft-wall. In fact, the problem reduces to examining the limit $\lambda\rightarrow 1$  while keeping $v$ fixed. A calculation along the lines of \cite{az1} gives the leading terms 

\begin{equation}\label{lE-large}
	\vert QY\vert=-\frac{1}{\sqrt{\s}}\ln(1-\lambda)+O(1)
	\,,\qquad
	E=-\g\ep\sqrt{\s}\,\ln(1-\lambda)+O(1)
	\,.
\end{equation}
Combining these equations results in 

\begin{equation}\label{E10}
	E=\sigma\vert QY\vert+O(1)
	\,.
\end{equation}
As expected, at leading order $E$ is linear in $\vert QY\vert$. The constant of proportionality $\sigma$ is called the string tension, given explicitly in Eq.\eqref{sigma}. It is universal and independent of $v$.  

To find the constant term in the asymptotic expansion of $E$ for large $\vert QY\vert$, consider

 \begin{equation}\label{E11}
	\begin{split}
	E-\sigma\vert QY\vert=\g &\sqrt{\frac{\s}{\lambda}}
	\biggl(
	\int_0^1\frac{du}{u^2}
\Bigl(\ep^{\lambda u^2}
\Bigl[1-\lambda u^4\ep^{1+\lambda(1-2u^2)}\Bigr]
\Bigl[1-u^4\ep^{2\lambda (1-u^2)}\Bigr]^{-\frac{1}{2}}
-1-u^2\Bigr)
\\
+&
\int_{\sqrt{\frac{v}{\lambda}}}^1\frac{du}{u^2}\ep^{\lambda u^2}
\Bigl[1-\lambda u^4\ep^{1+\lambda (1-2u^2)}\Bigr]
\Bigl[1-u^4\ep^{2\lambda (1-u^2)}\Bigr]^{-\frac{1}{2}}
\biggr)\,+c
	 \,.
	 \end{split}
\end{equation}

Letting $\lambda=1$, we get 

\begin{equation}\label{E12}
E-\sigma\vert QY\vert=\g\sqrt{\s}
	\biggl(
	\int_0^1\frac{du}{u^2}
\Bigl(\ep^{u^2}
\Bigl[1-u^4\ep^{2(1-u^2)}\Bigr]^{\frac{1}{2}}
-1-u^2\Bigr)
+
\int_{\sqrt{v}}^1\frac{du}{u^2}\ep^{u^2}
\Bigl[1-u^4\ep^{2(1-u^2)}\Bigr]^{\frac{1}{2}}
\biggr)\,+c+o(1)
	 \,.
\end{equation}
To this order, the expansion of $E$ is therefore 

\begin{equation}\label{E13}
	E=\sigma\vert QY\vert-\g\sqrt{\s}\,{\cal I}(v)
	+c+o(1)
	\,,
\end{equation}
where the function ${\cal I}$ is defined by Eq.\eqref{I}.

\begin{figure}[htbp]
\centering
\includegraphics[width=8.25cm]{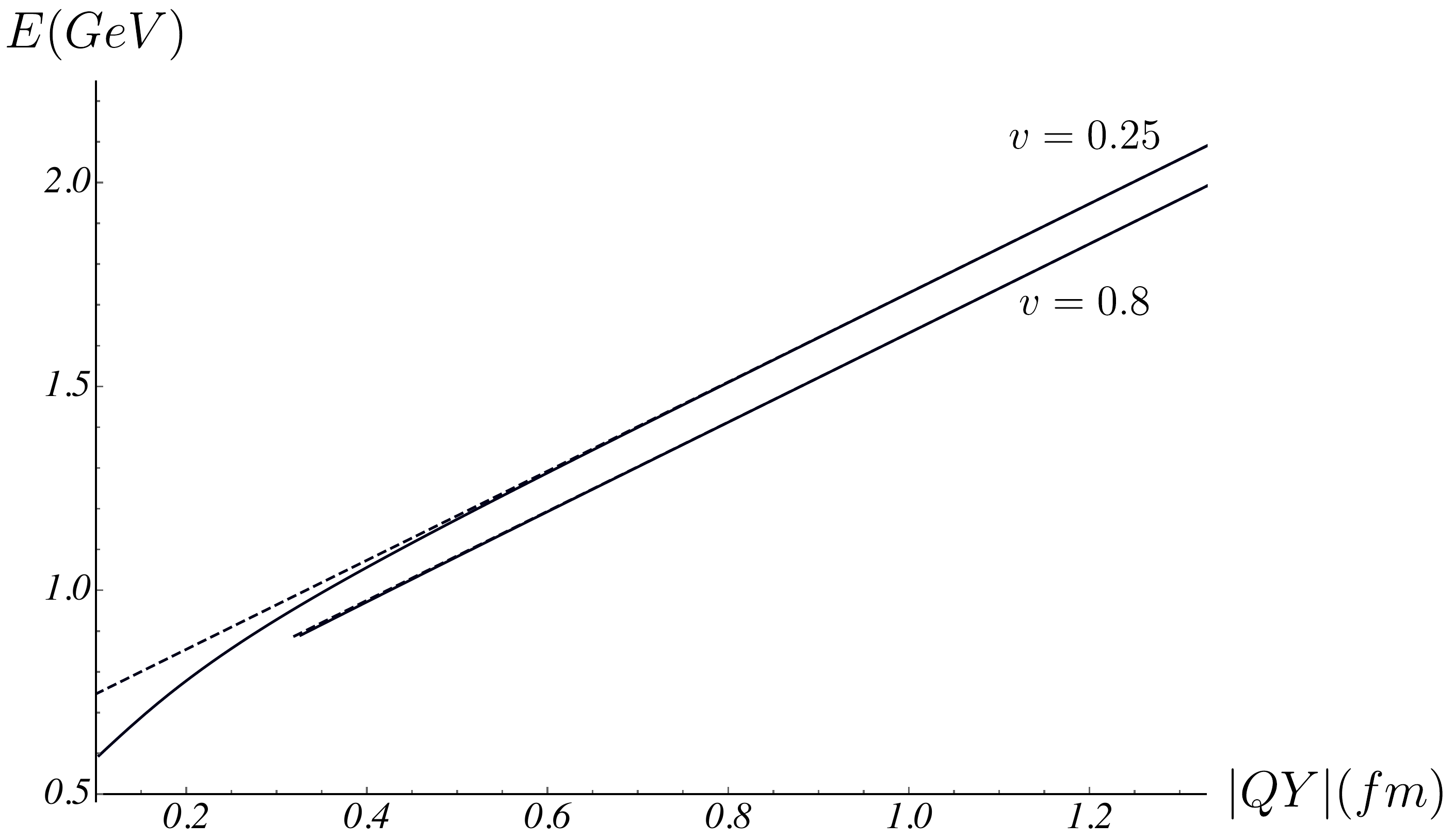}
\caption{{\small $E$ vs $\vert QY\vert$ for $\vert QY\vert\geq {\cal L}^-(v,v)/\sqrt{s}$ and fixed $v$. Here we use the $L$ set with $c=0.623\,\text{GeV}$. The dashed lines represent the asymptotic expression \eqref{E13}.}}
\label{EQY}
\end{figure}

It is instructive to see how good this linear approximation actually is. To this end, in Figure \ref{EQY} we plot both  
functions. We see that the linear approximation is perfect for strings longer that $0.6\,\text{fm}$. This fact enables one to find expressions for the string breaking distance analytically.
\section{Gluing conditions}
\renewcommand{\theequation}{C.\arabic{equation}}
\setcounter{equation}{0}

The string solutions we discussed in Appendix B are building blocks of multi-string configurations. What is still missing are gluing conditions for such blocks. Our goal here is to provide those conditions, with particular attention to the configurations of Sect. III and IV. More detail can be found, for example, in \cite{a-3q}. 

A general force balance equation at a baryon vertex, takes the form 

\begin{equation}\label{fbv-g}
	\mathbf{e}_1+\mathbf{e}_2+\mathbf{e}_3+\mathbf{f}_v=0
	\,. 
\end{equation}
Here $\mathbf{e}$'s are the string tensions and $\mathbf{f}_v$ is a gravitational force acting on the vertex. The presence of this force is the main difference between string models in flat space and those in curved spaces. In the model we are considering $\mathbf{f}_v$ has only one non-zero component which points in the $r$-direction. A formula for this component can be derived from the action $S_{\text{vert}}=T E_{\text{vert}}$. Explicitly, $\mathbf{f}_v^r=-\delta E_{\text{vert}}/\delta r=-\tau_v\partial_{r}\frac{\ep^{-2\s r^2}}{r}$, with $r$ the coordinate of the vertex. 

\begin{figure}[htbp]
\centering
\includegraphics[width=6.75cm]{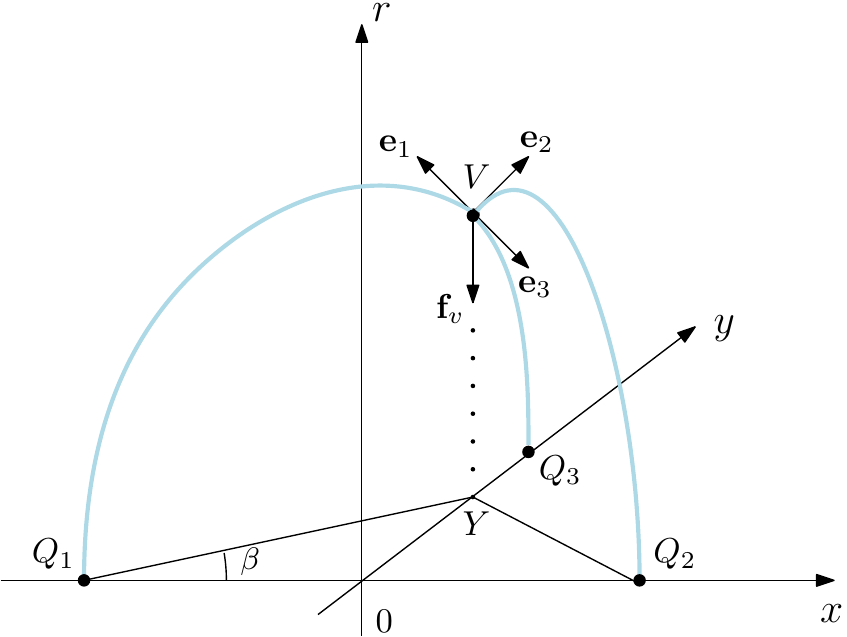}
\hspace{2cm}
\includegraphics[width=6.75cm]{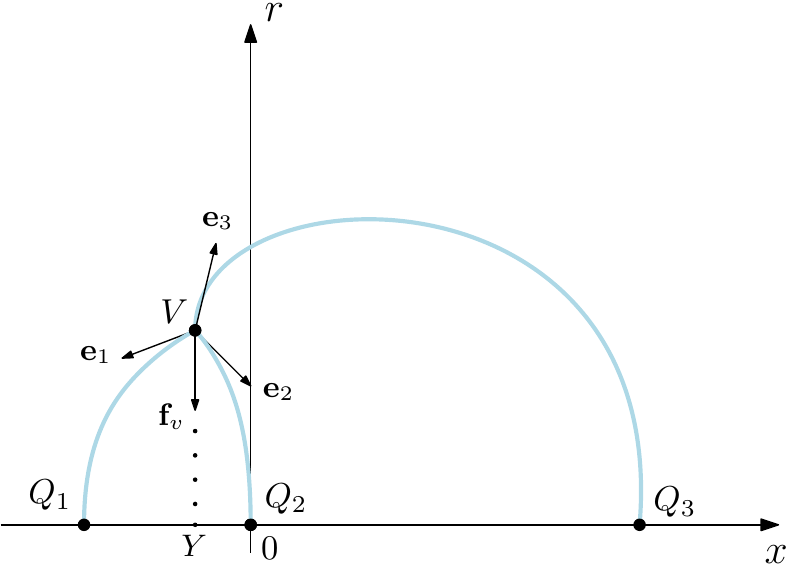}
\caption{{\small Three strings meeting at a baryon vertex in the bulk. The gravitational force $\mathbf{f}_v$ acting on the vertex is directed in the downward vertical direction. Left: The non-planar case with heavy quarks are at the vertices of an isosceles triangle. Right: The planar case with heavy quarks are on the $x$-axis.}}
\label{gluing}
\end{figure}

Now consider two important examples. The first example is illustrated in the left panel of Figure \ref{gluing}. 
Using the formula \eqref{e} for the string tension, it can be shown that in the present case the tensions are written in components as $\mathbf{e}_1=-\g w(\rv)(\cos\beta\cos\alpha_1,\sin\beta\cos\alpha_1,\sin\alpha_1)$, $\mathbf{e}_2=-\g w(\rv)(-\cos\beta\cos\alpha_1,\sin\beta\cos\alpha_1,\sin\alpha_1)$, and $\mathbf{e}_3=-\g w(\rv)(0,-\cos\alpha_3,\sin\alpha_3)$.\footnote{On symmetry grounds, the angles $\alpha_1$ and $\alpha_2$ are equal.} With this, the resulting equations for the $y$ and $r$ components take the form 

\begin{equation}\label{fbv}
\cos\alpha_3-2\sin\beta\cos\alpha_1=0
	\,,\qquad
2\sin\alpha_1+\sin\alpha_3-3\k(1+4v)\ep^{-3v}
=0
	\,.
\end{equation}
The equation for the $x$ component is trivially satisfied because of symmetry.

As a special case, consider $\alpha_3=\alpha_1$ which corresponds to an equilateral triangle. The force balance equation now has only one non-trivial component that gives rise to 

\begin{equation}\label{alpha-etg}
\sin\alpha_1-\k(1+4v)\ep^{-3v}
=0
\,.
\end{equation}

The second example is illustrated in the right panel of the Figure. The analysis proceeds in a similar manner as above. The string tensions are written in components as $\mathbf{e}_1=-\g w(\rv)(\cos\alpha_1,\sin\alpha_1)$, $\mathbf{e}_2=\g w(\rv)(\cos\alpha_2,-\sin\alpha_2)$, and $\mathbf{e}_3=\g w(\rv)(\cos\alpha_3,-\sin\alpha_3)$. So, in component form the force balance equation reads

 \begin{equation}\label{fbv-cg}
\cos\alpha_1-\cos\alpha_2-\cos\alpha_3=0
\,,\qquad
\sin\alpha_1+\sin\alpha_2+\sin\alpha_3=3\k(1+4v)\ep^{-3v}
\,.	
\end{equation}

\section{Some string configurations}
\renewcommand{\theequation}{D.\arabic{equation}}
\setcounter{equation}{0}

In order to make this paper more self-contained, in this Appendix we briefly review some basic results on the static string configurations with light quarks. For more details on these results, see \cite{a-strb,a-QQq}.

\subsection{The configurations for $Q\bar q$ and $Qqq$}

\begin{figure}[htbp]
\centering
\includegraphics[width=5cm]{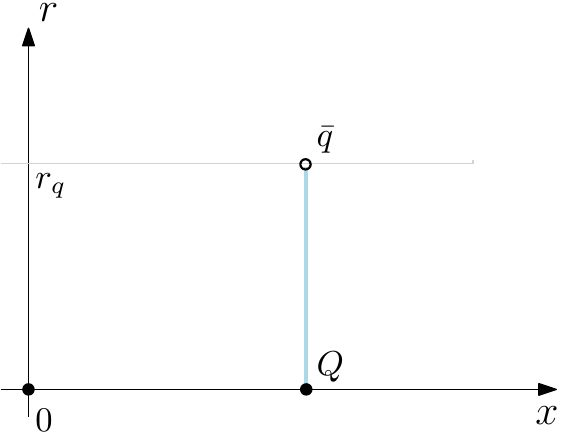}
\hspace{2.75cm}
\includegraphics[width=5cm]{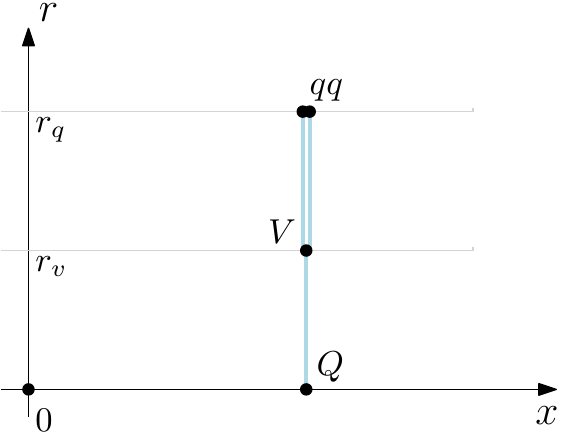}
\caption{{\small Left: A static configuration for a heavy-light meson $Q\bar q$. Right: A static configuration for a heavy-light baryon $Qqq$.}}
\label{Qqb}
\end{figure}

First, consider the configuration of Figure \ref{Qqb} on the left. In the context of the gauge/string duality it provides the description of a heavy-light meson in the static limit. 
The configuration is governed by the action which is the sum of the Nambu-Goto action $S_{\text{NG}}$ and the boundary term $S_{\text{q}}$, $S=S_{\text{NG}}+S_{\text{q}}$.

A short calculation performed in \cite{a-strb} shows that the energy $E$ is expressed in terms of $q$ by 

\begin{equation}\label{EQqb}
E_{\Qqb}=\g\sqrt{\s}\Bigl({\cal Q}(q)+\n\frac{\ep^{\oh q}}{\sqrt{q}}\,\Bigr)+c
\,.
\end{equation}
Here $c$ is a normalization constant. The function ${\cal Q}$ is as defined in Appendix A. $q$ is a solution to the equation 

\begin{equation}\label{q}
\ep^{\frac{q}{2}}+\n(q-1)=0
\,,
\end{equation}
which is obtained by varying the action with respect to $\rq$. A noteworthy fact is that such a solution only exists for $q<1$. The physical meaning of this equation is that it is nothing else but the force balance equation at the string endpoint which says that the net force acting on the light antiquark is zero.  

Now let us discuss the remaining configuration on the right, which describes a static heavy-light baryon. This configuration is governed by the following action $S=\sum_{i=1}^3S_{\text{NG}}^{(i)}+2S_{\text{q}}+S_{\text{vert}}$, which is the sum of the Nambu-Goto actions, boundary actions $S_{\text{q}}$, and action $S_{\text{vert}}$. 

The energy of this configuration can be written in the form \cite{a-strb}

\begin{equation}\label{EQqq}
	E_{\Qqq}=\g\sqrt{\s}\Bigl(2{\cal Q}(q)-{\cal Q}(\vp)
	+2\n\frac{\ep^{\oh q}}{\sqrt{q}}
	+3\k\frac{\ep^{-2\vp}}{\sqrt{\vp}}\,\Bigr)+c
	\,,
\end{equation}
where $\vp$ is a solution to  

\begin{equation}\label{v||}
1+3\k(1+4v)\ep^{-3v}=0
\,.
\end{equation}
The above equation defines the position of the baryon vertex in the bulk, whereas the 
equation \eqref{q} defines the position of the light quarks. At vanishing baryon chemical potential light quarks and antiquarks are at the same radial distance from the boundary. Note that the stable configuration exists only for $q\geq v$.
\subsection{The configurations for $QQq$}

\begin{figure}[htbp]
\centering
\includegraphics[width=5.6cm]{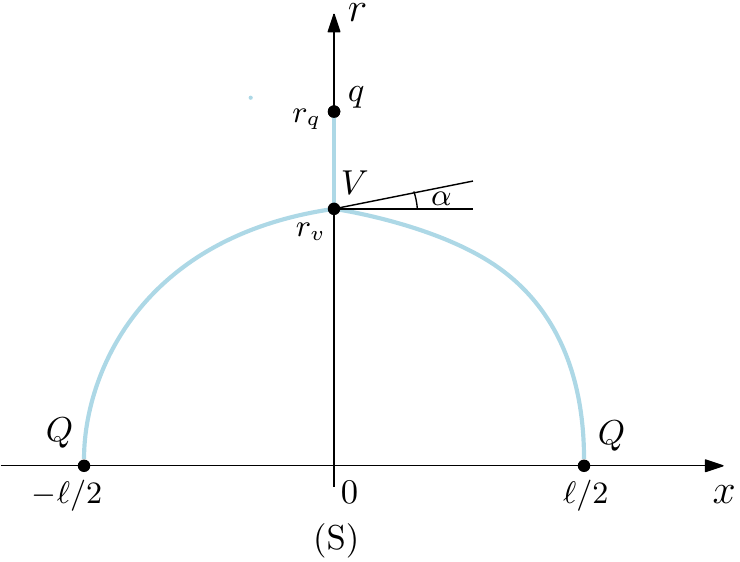}
\hspace{0.2cm}
\includegraphics[width=5.6cm]{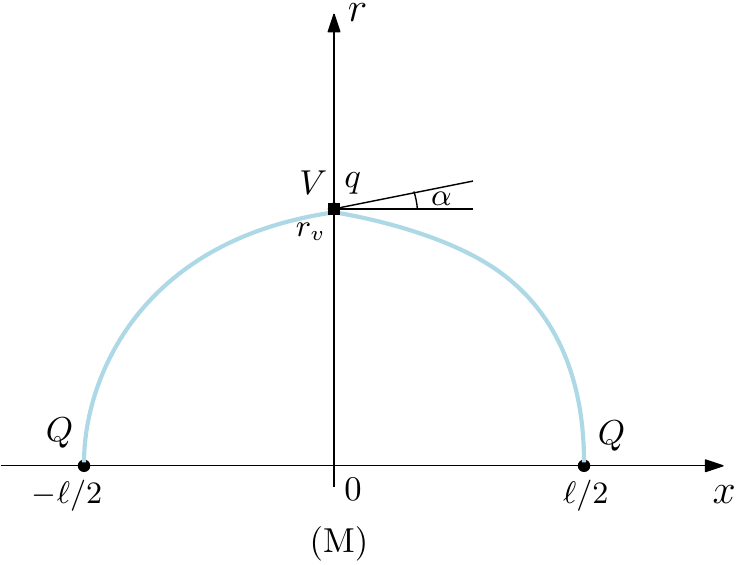}
\hspace{0.2cm}
\includegraphics[width=5.8cm]{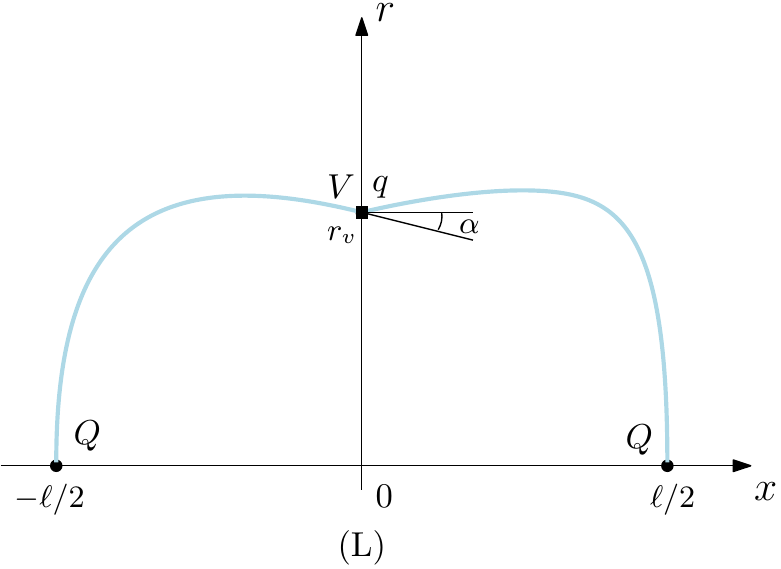}
\caption{{\small Three types of static string configurations that contribute to the ground state of the $QQq$ system. $\alpha$ denotes a tangent angle of the left string at point $V$.}}
\label{QQq}
\end{figure}

Finally, we consider the string configurations of Figure \ref{QQq}. 
Those configurations were proposed to model the $QQq$ system in \cite{a-QQq}, whose conventions we follow here. The important point is that the shape of the configuration changes with the increase of heavy quark separation. 

For small $\ell$, the corresponding configuration is labeled by (S). In this case, the relation between the energy and heavy quark separation is written in parametric form  

\begin{equation}\label{EQQqs}
\ell^{\Se}=\frac{2}{\sqrt{\s}}{\cal L}^+(\alpha,v)
\,,
\qquad
E_{\QQq}^{\Se}=\g\sqrt{\s}
\Bigl(
2{\cal E}^+(\alpha,v)
+
\n\frac{\ep^{\oh q}}{\sqrt{q}}+3\k\frac{\ep^{-2v}}{\sqrt{v}}
+
{\cal Q}(q)-{\cal Q}(v)
\Bigr)
+2c\,,
\end{equation}
with the parameter $v$ varying from $0$ to $q$. The value of $q$ is determined from equation \eqref{q}. The functions ${\cal L}^+$ and ${\cal E}^+$ are as defined in Appendix A. $c$ is a normalization constant. The tangent angle  $\alpha$ can be expressed in terms of $v$ by using the force balance equation at the vertex, with the result

\begin{equation}\label{alpha1}
\sin\alpha=\oh\bigl(1+3\k(1+4v)\ep^{-3v}\bigr)
\,.	
\end{equation}

At this point, it is worth mentioning that in the limit $\ell\rightarrow 0$ the energy reduces to a sum of energies: $E_{\QQq}^{\Se}=E_{\qQb}+E_{\QQ}$, as expected from heavy quark-diquark symmetry. This limit corresponds to small values of $v$, where the function $\ell(v)$ behaves exactly as described by Eq.\eqref{l-v}. 

For intermediate values of $\ell$, the configuration is labeled by (M). It looks like one of the strings is missing so that the position of the light quark coincides with that of the vertex. So, one has $q=v$. The distance $\ell$ is expressed in terms of $v$ and $\alpha$ by the same formula as before, only for another parameter range, whereas the energy by 

\begin{equation}\label{EQQqm} 
E_{\QQq}^{\Me}=\g\sqrt{\s}
\Bigl(
2{\cal E}^+(\alpha,v)
+
\frac{1}{{\sqrt{v}}}\bigl(
\n\ep^{\oh v}+
3\k\ep^{-2v}
\bigr)
\Bigr)
+2c\,.
\end{equation}
Now $v$ varies from $q$ to $\vm$, where $\vm$ is a solution to 

\begin{equation}\label{v0}
\n (1-v)+3\k(1+4v)\ep^{-\frac{5}{2}v}=0
\,.
\end{equation}
The force balance equation in this case gives 

\begin{equation}\label{alpha2}
\sin\alpha=\oh\bigl(
\n (1-v)\ep^{-\oh v}+3\k(1+4v)\ep^{-3v}
\bigr)
\,.
\end{equation}
A noteworthy fact is that $\alpha(\vm)=0$. 

For large $\ell$, the proper configuration is that labeled by (L). In fact, what happens in the transition from (M) to (L) is that the tangent angle changes the sign from positive to negative. Keeping this in mind makes it much easier to arrive at the relation between the energy and quark separation. Replacing ${\cal L}^+$ and ${\cal E}^+$ by ${\cal L}^-$ and ${\cal E}^-$, it becomes

\begin{equation}\label{EQQql}
\ell^{\Le}=\frac{2}{\sqrt{\s}}
{\cal L}^-(\lambda,v)
\,,
\qquad
E_{\QQq}^{\Le}=\g\sqrt{\s}
\Bigl(
2{\cal E}^-(\lambda,v)
+
\frac{1}{\sqrt{v}}
\bigl(\n\ep^{\oh v}+3\k\ep^{-2v}\bigr)
\Bigr)
+2c\,,
\end{equation}
with the parameter $v$ varying from $\vm$ to $\vl$. The upper bound is found by solving the non-linear equation 

\begin{equation}\label{v1}
2\sqrt{1-v^2\ep^{2(1-v)}}+3\k(1+4v)\ep^{-3v}+\n (1-v)\ep^{-\oh v}=0
\,
\end{equation}
on the interval $[0,1]$. A function $\lambda(v)$ is written explicitly as 

\begin{equation}\label{lambda-QQq}
\lambda(v)=-\text{ProductLog}\Bigl[-v\ep^{-v}
\Bigl(1-\frac{1}{4}\Bigl(3\k(1+4v)\ep^{-3v}
+
\n(1-v)\ep^{-\oh v}\Bigr)^2
\,\Bigr)^{-\frac{1}{2}}
\,\Bigl]
\,,
\end{equation}
as it follows from Eq.\eqref{v-lambda} with the tangent angle found from the force balance equation at $V$. Note that $\lambda(\vl)=1$ that corresponds to the limit of infinitely long strings. 

One can summarize all this by saying that the energy as a function of the heavy quark separation is given in parametrical form by the two piecewise functions $E_{\QQq}=E_{\QQq}(v)$ and $\ell=\ell(v)$.

For future reference, it is worth noting that the asymptotic behavior of $E_{\QQq}(\ell)$ for large $\ell$ is 

\begin{equation}\label{EQQq-large}
	E_{\QQq}=\sigma\ell-2\g\sqrt{\s}I_{\QQq}+2c+o(1)
	\,,
	\end{equation}
with 
\begin{equation}\label{IQQq}
	I_{\QQq}={\cal I}(\vl)
-
\frac{1}{2\sqrt{\vl}}\Bigl(\n\ep^{\oh\vl}+3\k\ep^{-2\vl}\Bigr)
\,
\end{equation}
and the same string tension $\sigma$ as in \eqref{E13}.



\small
\begin{thebibliography}{99}
\bibitem{bj}
J. D. Bjorken, Is the $ccc$ a New  Deal for Baryon Spectroscopy?, Report No. FERMILAB-Conf-85/69, 1985.
\bibitem{richard}
J.-M. Richard, Phys.Rep. {\bf 212}, 1 (1992); Few Body Syst. {\bf 61}, 4 (2020).
\bibitem{bali}
G.S. Bali, Phys.Rep. {\bf 343}, 1 (2001).
\bibitem{3Q-lattice}
The following is an incomplete list: 
T.T. Takahashi, H. Suganuma, Y. Nemoto, and H. Matsufuru, Phys.Rev.D {\bf 65}, 114509 (2002); C. Alexandrou, Ph. de Forcrand, and O. Jahn, Nucl. Phys.Proc.Suppl. {\bf 119}, 667 (2003); V. G. Bornyakov {\it et al}. [DIK Collaboration], Phys. Rev.D {\bf 70}, 054506 (2004); F. Bissey, F-G. Cao, A.R. Kitson, A.I. Signal, and D.B. Leinweber, Phys.Rev.D {\bf 76}, 114512 (2007); N. Sakumichi and H. Suganuma, Phys.Rev.D {\bf 92} 034511 (2015); Y. Koma and M. Koma, Phys.Rev.D {\bf 95}, 094513 (2017).
\bibitem{stringP}
X. Artru, Nucl. Phys. {\bf B85}, 442 (1975); G.C. Rossi and G. Veneziano, Nucl.Phys.B {\bf 123}, 507 (1977); N. Isgur and J. Paton, Phys.Rev.D {\bf 31}, 2910 (1985).
\bibitem{bulava}
J. Bulava, B. H\"orz, F. Knechtli, V. Koch, G. Moir, C. Morningstar, and M. Peardon, Phys.Lett.B {\bf 793}, 493 (2019).
\bibitem{drum}
I.T. Drummond, Phys.Lett.B {\bf 434}, 92 (1998).
\bibitem{a-3q} 
 O. Andreev, Phys.Lett.B {\bf 756}, 6 (2016); Phys.Rev.D {\bf 93}, 105014 (2016).
\bibitem{a-strb}
O. Andreev, Phys.Lett.B {\bf 804}, 135406 (2020); Phys.Rev.D {\bf 101}, 106003 (2020).
\bibitem{az1}
O. Andreev and V.I. Zakharov, Phys.Rev.D {\bf 74}, 025023 (2006).
\bibitem{son}
J. Erlich, E. Katz, D.T. Son, and M.A. Stephanov, Phys.Rev.Lett. {\bf 95}, 261602 (2005).
\bibitem{witten}
E. Witten, J. High Energy Phys. {\bf 9807}, 006 (1998).
\bibitem{wolf}
See, e.g., MathWorld - A Wolfram Web Resource. https://reference.wolfram.com/language/ref/ProductLog.html.
\bibitem{a-3q0}
O. Andreev, Phys.Rev.D {\bf 78}, 065007 (2008).
\bibitem{a-q2}
O. Andreev, Phys.Rev.D {\bf 73}, 107901 (2006).
\bibitem{lipkin}
O.W. Greenberg and H.J. Lipkin, Nucl.Phys.A {\bf 370}, 349 (1981).
\bibitem{PC}
M.V. Carlucci, F. Giannuzzi, G. Nardulli, M. Pellicoro, and S. Stramaglia, Eur.Phys.J. {\bf C57}, 569 (2008).
\bibitem{wise}
M.J. Savage and M.B. Wise, Phys.Lett.B {\bf 248}, 177 (1990).
\bibitem{a-QQq}
O. Andreev, J. High Energy Phys. {\bf 05}, 173 (2021).
\bibitem{amp}
A.M. Polyakov, Confinement and liberation, hep-th/0407209. 
\bibitem{malda}
J.M. Maldacena, Phys.Rev.Lett. {\bf 80}, 4859 (1998).



\end{thebibliography}
\end{document}